\documentclass[12pt]{article}
\usepackage{epsfig}
\begin{document}
\title{Medium effects in ${\Lambda}K^+$ pair production\\
               by 2.83 GeV protons on nuclei}
\author{E. Ya. Paryev$^{1,2}$, M. Hartmann$^3$, Yu. T. Kiselev$^2$\\
{\it $^1$Institute for Nuclear Research, Russian Academy of Sciences,}\\
{\it Moscow 117312, Russia}\\
{\it $^2$Institute for Theoretical and Experimental Physics,}\\
{\it Moscow 117218, Russia}\\
{\it $^3$Institut f$\ddot{u}$r Kernphysik and J$\ddot{u}$lich Centre for Hadron Physics,}\\
{\it Forschungszentrum J$\ddot{u}$lich, D--52425 J$\ddot{u}$lich, Germany}}

\renewcommand{\today}{}
\maketitle

\begin{abstract}
   We study ${\Lambda}K^+$ pair production in the interaction of protons of 2.83 GeV kinetic energy with C, Cu, Ag, and Au target nuclei
   in the framework of the nuclear spectral function approach for incoherent primary proton--nucleon and
   secondary pion--nucleon production processes, and processes associated with the creation of intermediate
   ${\Sigma^0}K^+$ pairs.
   The approach accounts for the initial proton and final $\Lambda$ hyperon absorption, final $K^+$ meson distortion in nuclei, target nucleon binding, and Fermi motion, as well as nuclear mean-field potential effects on these processes.
   We calculate the $\Lambda$ momentum dependence of the absolute ${\Lambda}K^+$ yield from the  target nuclei considered, in the kinematical conditions of the ANKE experiment, performed at COSY, within the different scenarios for the $\Lambda$-nucleus effective scalar potential.
   We show that the above observable is appreciably sensitive to this potential in the low-momentum
   region.
   Therefore, direct comparison of the results of our calculations with the data from the ANKE-at-COSY experiment can help to determine the above potential at finite momenta.
   We also demonstrate that the two-step pion--nucleon production channels dominate in the low-momentum
   ${\Lambda}K^+$ production in the chosen kinematics and, therefore, they have to be taken into account
   in the analysis of these data.
\end{abstract}

\newpage

\section*{1 Introduction}
\hspace{1.5cm}The production of strangeness carrying particles has been intensively investigated for a long time in many experiments, using a variety of beams, targets and energies. Strange particles produced in proton-nucleus and nucleus-nucleus reactions interact with the hadronic environment not only by collisions but also by potential interaction, which is expected to lead to a  change in the particle properties in matter. Numerous experiments performed during the past two decades were aimed at studying the properties of kaons, antikaons and hyperons in a strongly interacting matter. They were motivated by the theoretically predicted phenomena of the partial restoration of chiral symmetry in hot/dense nuclear matter and the possible existence of an antikaon condensate or the presence of hyperons in the dense core of neutron stars (see Ref.~[1] for a recent review).

	The in-medium modification effects on the strange meson properties, mass and width, at nuclear saturation density can be described in terms of complex nuclear optical potential. It has been established that the real parts of the $K^+$ and $K^0$ repulsive nuclear potentials amount to 20--40 MeV at normal nuclear density $\rho_0=0.16$ fm$^{-3}$ [2], while the attractive $K^-$ potential is stronger, although there is no common agreement about its strength so far. The calculations, based on chiral Lagrangians [3,4,5] or on meson-exchange potentials [6], predict a relatively shallow low-energy $K^-$--nucleus potential with central depth of the order of -50 to -80 MeV. On the other hand, fits to the $K^-$ atomic data in terms of phenomenological density-dependent optical potential or relativistic mean-field calculations [7] lead to a much stronger potential with depth of about -200 MeV at density $\rho_0$. The reported results of different measurements indicate that the values of the central depth of the potential are spread out over a wide range from -30 to -200 MeV [8]. The recent data for kaon pair production by protons off nuclei obtained by the ANKE Collaboration do not favor a deep antikaon potential [9]. The imaginary part of the nuclear optical potential, which is responsible for the absorption of strange mesons with both open ($K^+$, $K^-$) and hidden ($\phi$) strangeness during their way out of nuclei, has been also studied by measuring the so-called transparency ratio [10]. These investigations provide information on the in-medium meson width or on the in-medium meson--nucleon cross section.

	Essential progress has been made over past decades in studying the properties of strange mesons in nuclear matter produced in proton--nucleus and nucleus--nucleus collisions at intermediate energies within the different
transport models (QMD, IQMD and HSD, GiBUU and LQMD) as well as in investigating in these collisions,
by means of kaons, the high-density behavior of the EoS (see Ref.~[11] for a review and Ref.~[12] for a recent study). It has been  shown that the kaon and antikaon mean-field potentials change the structure of the majority
of the observables (collective flows, inclusive spectra, their ratios) and make it possible to get good agreement
with the available experimental data. Much less attention, however, has been paid to the properties of strange baryons in the surrounding nuclear environment in these collisions at near-threshold beam energies.
Recently, $\Lambda$ and $\Sigma^{+,0,-}$ production in heavy-ion- and proton-induced reactions on nuclei near
threshold energies has been investigated within the isospin- and momentum-dependent LQMD transport model [12].
In particular, it was found that the $\Sigma^-/\Sigma^+$ ratio is sensitive to the isospin asymmetric part of the
EoS, which is poorly understood up to now. Knowledge of this part is important in astrophysics. The influence
of the lambda potential on inclusive $\Lambda$ creation in $pA$ interactions, however, turned out to be negligible.
A large series of results on $\Lambda$ and $\Sigma$ hyperon production in proton--nucleus collisions has been compiled in the high energy region from 9 to 400 GeV [13]. Experimental information on their creation at lower proton beam energies is very poor, although the search for medium modification effects here, in particular, in the near-threshold energy domain, looks quite promising. So far, the only experiment here aimed at the study of $\Lambda(1115)$ hyperon
production in $p+$Nb reactions at proton beam energy of 3.5 GeV has been performed by the HADES Collaboration at SIS18/GSI Darmstadt [14]. This kinetic energy corresponds to an excess energy of 0.63 GeV with respect to the threshold of $\Lambda(1115)$ production in $NN$ collisions  and to a typical hyperon momentum of about 0.4 GeV/c. However, investigation of  possible medium effects has been out of the scope of the performed data analysis, which was directed at studying the most important features of hyperon production.
Rich hypernuclear experimental information [15] enables us to study
the hyperon--nucleus interaction at (almost) zero hyperon momentum with respect to the nuclear matter.

	The in-medium properties of hyperons at finite momentum, density and temperature have become a matter of intense theoretical investigation over  past years. Thus, the medium modification of the
$\Lambda(1520)$ hyperon has been studied within chiral unitary theory with coupled channels [16]. It was found that at normal nuclear matter density the mass shift of the $\Lambda(1520)$ is small (about 2\%), while its in-medium width is more than five times bigger than the free one. The impact of the in-medium $\Lambda(1520)$ width on the hyperon yield from photon-- and proton--nucleus reactions has been analyzed in the framework of a collision model based on eikonal approximation in Ref.~[17]. The spectral functions for the hyperons have been evaluated in a self-consistent and covariant many-body approach [18]. Attractive mass shifts of about -30 and -40 MeV/c$^2$ have been predicted for the $\Lambda(1405)$ and $\Sigma(1385)$ hyperons,  respectively, at rest
at normal nuclear matter density. It has also been found that the $\Lambda(1115)$ hyperon downward mass shift of about -30 MeV/c$^2$ at this density is quite independent of the three-momentum, but its in-medium width is significantly increased as the hyperon moves with respect to the bulk matter. The predicted mass shift for the $\Sigma(1195)$ is about -22 MeV/c$^2$ at saturation density.
The in-medium properties of the $\Lambda(1115)$, $\Sigma(1195)$ and $\Sigma(1385)$ hyperons have been investigated in the chiral unitary approach [19]. It has been found that $\Lambda(1115)$,
$\Sigma(1195)$ and $\Sigma(1385)$ experience an attractive potential of about -50, -40 and -10 MeV, respectively, at normal nuclear matter density and at zero momentum relative to the surrounding
nuclear matter. The imaginary parts of the optical potentials of $\Lambda(1115)$ and
$\Sigma(1195)$ hyperons are less in absolute magnitude than 10 MeV, while that for the $\Sigma(1385)$ hyperon
amounts to -50 MeV at this density and momentum. In contrast to Ref.~[18], both the real and imaginary parts of the optical potential for all considered hyperons change significantly  in the momentum range from 0 to 600 MeV/c. Recently the momentum dependences of the real and imaginary parts of the single-particle potential of the $\Lambda(1115)$ and $\Sigma(1195)$ hyperons in isospin symmetric and asymmetric nuclear matter have been studied in chiral effective field theory [20]. At saturation density, the real part of the attractive $\Lambda(1115)$ potential is between -(24-28) MeV at zero momentum with respect to the nuclear matter, increases with the hyperon momentum and becomes positive (repulsive)  at momenta exceeding 400 MeV/c, while a $\Sigma(1195)$ feels a repulsive potential even at zero momentum. Cited in Refs.~[18-20], the values of the real part of a $\Lambda(1115)$ potential at saturation density and at zero momentum in the nuclear
matter rest frame are somewhat different, but comparable to an empirical value of about -30 MeV deduced from the analysis of binding energies of hypernuclei [15]. On the other hand, the situation with, in particular, the
$\Lambda$--nuclear potential at finite momenta, is still unclear at present, in spite of a lot theoretical activity in this field. There is currently no experimental information about the momentum dependence of this potential. This information can be deduced from analysis of the experimental data on the production of $\Lambda$ hyperons in coincidence with $K^+$ mesons in proton collisions with
C, Cu, Ag and Au targets at an initial energy of 2.83 GeV, taken recently by the ANKE Collaboration at the COSY
accelerator. The advantage of such coincident data compared to inclusive data is that fewer individual exclusive elementary $\Lambda$ production channels need to be accounted for in their
interpretation, with the aim of obtaining information on the $\Lambda$--nuclear potential, compared to what is involved in the analysis of inclusive data. That makes such interpretations clearer and allows for reduction of the theoretical uncertainties associated with the $\Lambda$ particle production mechanisms.
Moreover, one may hope that such (more differential than the inclusive) coincident data
in combination with the available high-energy $pA$ data will allow us to investigate the $\Lambda$-nuclear potential.

   In this connection, the main goal of the present work is to give predictions for the absolute yields
of ${\Lambda}K^+$ pairs from $pA$ collisions in the kinematical conditions of the ANKE experiment, adopting
the collision  model, based on the nuclear spectral function, for incoherent one-step and two-step ${\Lambda}K^+$ pair creation processes in different scenarios for the lambda--nuclear potential.
Direct comparison of these predictions with the expected data from this experiment will allow us
to shed light on the $\Lambda$ potential in a nuclear medium at finite momenta.

\section*{2 The model}
\section*{2.1 Direct ${\Lambda}K^+$ production mechanisms}

\hspace{1.5cm}The direct production of $\Lambda$ hyperons in coincidence with forward going $K^+$
mesons in the kinematical conditions of the ANKE experiment
in $pA$ collisions at incident energy of 2.83 GeV of our interest can occur in the following
$pp$ and $pn$ elementary processes with zero, one and two pions in the final states
\footnote{$^)$Recall that the free threshold energies, e.g., for the processes $pp \to {\Lambda}pK^+$,
$pp \to {\Lambda}p{\pi^0}K^+$ and $pp \to {\Lambda}p{\pi^0}{\pi^0}K^+$ amount, respectively, to 1.58,
1.96 and 2.35 GeV. We can neglect at the beam energy of interest the subprocesses $pN \to {\Lambda}N3{\pi}K$
with three pions in the final states, due to the proximity of their production thresholds in free $pN$
interactions to this energy. Thus, for instance, the threshold energy of the channel
$pp \to {\Lambda}p3{\pi^0}K^+$ is 2.77 GeV. This energy is close to the incident proton energy of 2.83 GeV.
Hence, the subprocesses $pN \to {\Lambda}N3{\pi}K$ are energetically suppressed.}$^)$
:
\begin{eqnarray}
p+p \to \Lambda+p+K^+,
\end{eqnarray}
\begin{eqnarray}
p+p \to \Lambda+p+\pi^0+K^+,\nonumber\\
p+p \to \Lambda+n+\pi^++K^+;
\end{eqnarray}
\begin{eqnarray}
p+p \to \Lambda+p+\pi^0+\pi^0+K^+,\nonumber\\
p+p \to \Lambda+p+\pi^++\pi^-+K^+,\nonumber\\
p+p \to \Lambda+n+\pi^0+\pi^++K^+;
\end{eqnarray}
\begin{eqnarray}
p+n \to \Lambda+n+K^+,
\end{eqnarray}
\begin{eqnarray}
p+n \to \Lambda+n+\pi^0+K^+,\nonumber\\
p+n \to \Lambda+p+\pi^-+K^+;
\end{eqnarray}
\begin{eqnarray}
p+n \to \Lambda+n+\pi^0+\pi^0+K^+,\nonumber\\
p+n \to \Lambda+n+\pi^++\pi^-+K^+,\nonumber\\
p+n \to \Lambda+p+\pi^-+\pi^0+K^+.
\end{eqnarray}

  Let us now discuss the total cross sections of the reactions (1)--(6), which we will use throughout our
calculations of $\Lambda$$K^+$ pair yields in $pA$ collisions. The channel $pp \to {\Lambda}pK^+$ has been
extensively studied experimentally both earlier -- mostly at beam energies $\ge$ 2.85 GeV [21] -- and
recently -- at initial proton energies $\le$ 2.5 GeV by the COSY-11, COSY-TOF and ANKE Collaborations
(see, for example, [22--29]) as well as very recently--at incident proton energy of 3.5 GeV by the
HADES Collaboration [30].
We will use the following parametrization of available experimental data for the total
cross section of the $pp \to {\Lambda}pK^+$ reaction:
\begin{equation}
\sigma_{pp \to {\Lambda}pK^+}(\sqrt{s},\sqrt{s_{{\rm th}}})=\left\{
\begin{array}{ll}
	\frac{A_{\Lambda}(s-s_{{\rm th}})^2}{4m_p^2+B_{\Lambda}(s-s_{{\rm th}})^2}
	&\mbox{for $0.435~{\rm GeV} < \sqrt{s}-\sqrt{s_{{\rm th}}} < 2.0~{\rm GeV}$}, \\
	&\\
       \frac{C_{\Lambda}\left(\sqrt{s}-\sqrt{s_{{\rm th}}}\right)^2}
       {\left[1+\sqrt{1+(\sqrt{s}-\sqrt{s_{{\rm th}}})/D_{\Lambda}}\right]^2}
	&\mbox{for $0 < \sqrt{s}-\sqrt{s_{{\rm th}}} \le 0.435~{\rm GeV}$},
\end{array}
\right.	
\end{equation}
where $\sqrt{s}$ is the total $pp$ center-of-mass energy and
$\sqrt{s_{{\rm th}}}=m_{\Lambda}+m_p+m_{K^+}$ is the threshold energy;
$m_{\Lambda}$, $m_p$ and $m_{K^+}$ are the $\Lambda$ hyperon, proton and $K^+$
meson free space masses, respectively; and the constants
$A_{\Lambda}$, $B_{\Lambda}$, $C_{\Lambda}$ and $D_{\Lambda}$ are given as:
\begin{equation}
A_{\Lambda}=122.943~{\rm {\mu}b}/{\rm GeV}^2,\,\,\, B_{\Lambda}=2.015/{\rm GeV}^2,\,\,\,
C_{\Lambda}=25740~{\rm {\mu}b}/{\rm GeV}^2,\,\,\, D_{\Lambda}=5.203\cdot10^{-3}~{\rm GeV}.
\end{equation}
This combines the relevant fit from Ref.~[31] in the high excess energy region
with those given in Refs.~[28, 32] in the low energy interval.
In Fig. 1 the results of calculations by parametrization (7), (8)
for the total cross section $\sigma_{pp \to {\Lambda}pK^+}$ of reaction $pp \to {\Lambda}pK^+$ are shown as a solid line together with the available data [21--30] on this cross section in the considered range of excess energies.
The arrow on this figure indicates the excess
energy that corresponds to the beam kinetic energy of 2.83 GeV. This parametrization describes
the energy dependence of the cross section $\sigma_{pp \to {\Lambda}pK^+}$ quite well both at low and high
 excess energies considered.
\begin{figure}[!h]
\begin{center}
\includegraphics[width=12.0cm]{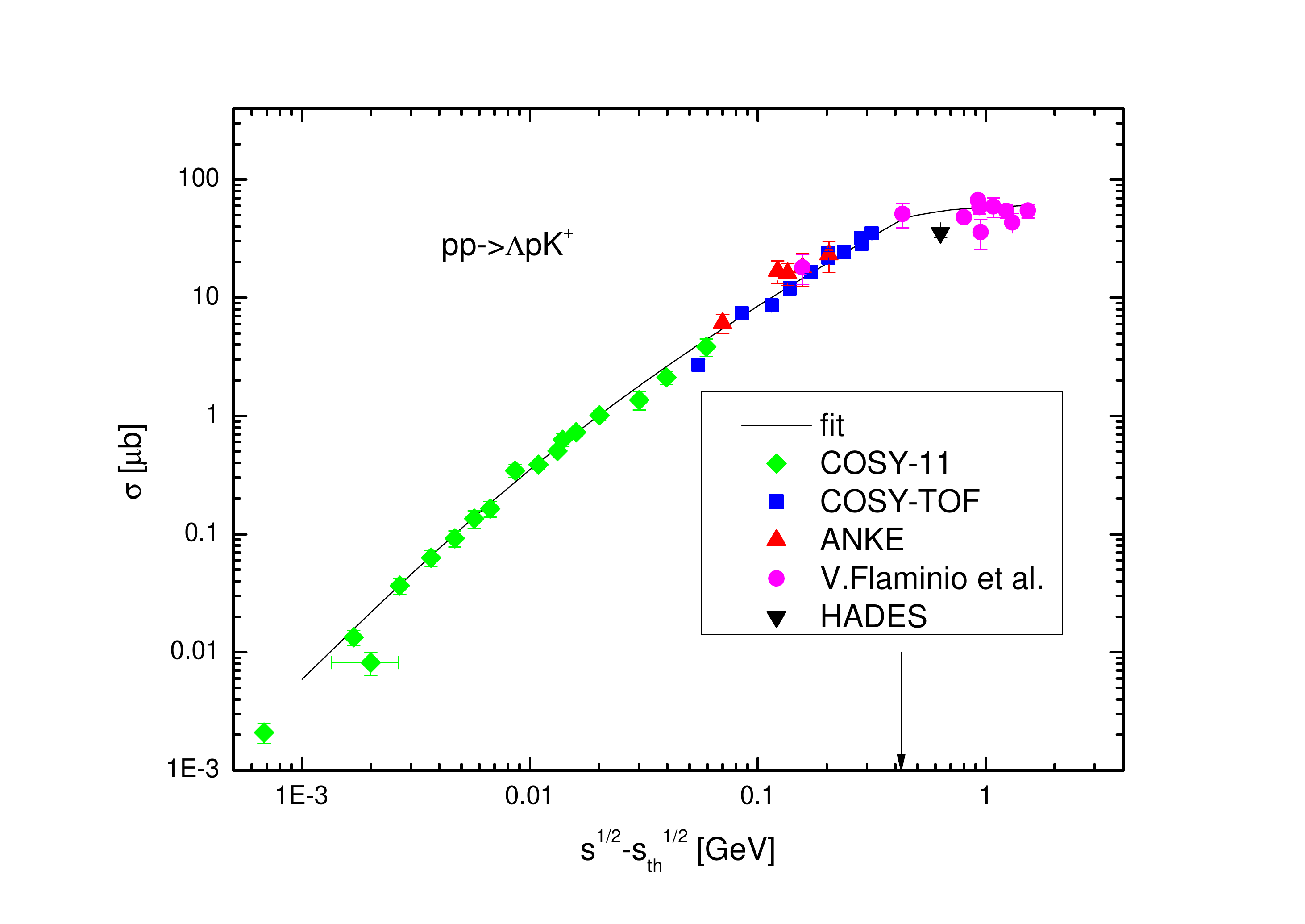}
\vspace*{-2mm} \caption{Total cross section for $pp \to {\Lambda}pK^+$ reaction
as a function of excess energy. For notation see the text.}
\label{void}
\end{center}
\end{figure}
Two processes (2) for $\Lambda$ production together with one pion in $pp$ interactions as well as the
process $pp \to {\Lambda}p{\pi^+}K^0$ have been measured at proton energies $\ge$ 4.1 GeV [21],
with an exception of two $pp \to {\Lambda}p{\pi^+}K^0$ measurements carried out by the ANKE [33] and HADES [34] Collaborations at 2.83 and 3.5 GeV, respectively. The three sets of data, which are available [21] at beam energies $\ge$ 4.1 GeV for these processes, indicate that they have similar total cross sections (see also Ref.~[35]):
\begin{equation}
\sigma_{pp \to {\Lambda}p{\pi^0}K^+} \approx \sigma_{pp \to {\Lambda}n{\pi^+}K^+} \approx
\sigma_{pp \to {\Lambda}p{\pi^+}K^0}.
\end{equation}
We assume that the relations (9) among the cross sections $\sigma_{pp \to {\Lambda}N{\pi}K}$ are also valid
at lower incident proton energies. For the free total cross section
$\sigma_{pp \to {\Lambda}p{\pi^+}K^0}$ we have used the following parametrization:
\begin{equation}
\sigma_{pp \to {\Lambda}p{\pi^+}K^0}({\sqrt{s},\sqrt{s_{1{\rm th}}}})=\left\{
\begin{array}{ll}
	1770.5\left(\sqrt{s}-\sqrt{s_{1{\rm th}}}\right)^{5.62}~[{\rm {\mu}b}]
	&\mbox{for $0<\sqrt{s}-\sqrt{s_{1{\rm th}}}\le 0.5~{\rm GeV}$}, \\
	&\\
                   72\left(\sqrt{s}-\sqrt{s_{1{\rm th}}}\right)~[{\rm {\mu}b}]
	&\mbox{for $0.5~{\rm GeV} < \sqrt{s}-\sqrt{s_{1{\rm th}}} < 3.0~{\rm GeV}$},
\end{array}
\right.	
\end{equation}
where $\sqrt{s_{1{\rm th}}}=m_{\Lambda}+m_p+m_{\pi^+}+m_{K^0}$ is the threshold energy.
Here, $m_{\pi^+}$ and $m_{K^0}$ are the $\pi^+$ and $K^0$ meson free space masses, respectively.
A comparison of the results of calculations by (10) (solid line)
with the experimental data for the $pp \to {\Lambda}p{\pi^+}K^0$ reaction from  ANKE [33]
(full triangle), from the HADES Collaboration [34] (full square
\footnote{$^)$This data point has been inferred from the measured total cross sections for direct and
resonant (via the intermediate $\Delta^{++}$) production of the 4-body final state ${\Lambda}p{\pi^+}K^0$
in $pp$ interactions.}$^)$
),
and for the data [21] at higher energies (full circles) is shown in Fig. 2.
In this figure we also show the predictions from the parametrization (45) (dashed line)
employed in the study [35] of kaon creation in heavy--ion collisions. This
parametrization significantly overestimates the lowest data point, obtained at 2.83 GeV initial proton kinetic energy.
\begin{figure}[!h]
\begin{center}
\includegraphics[width=12.0cm]{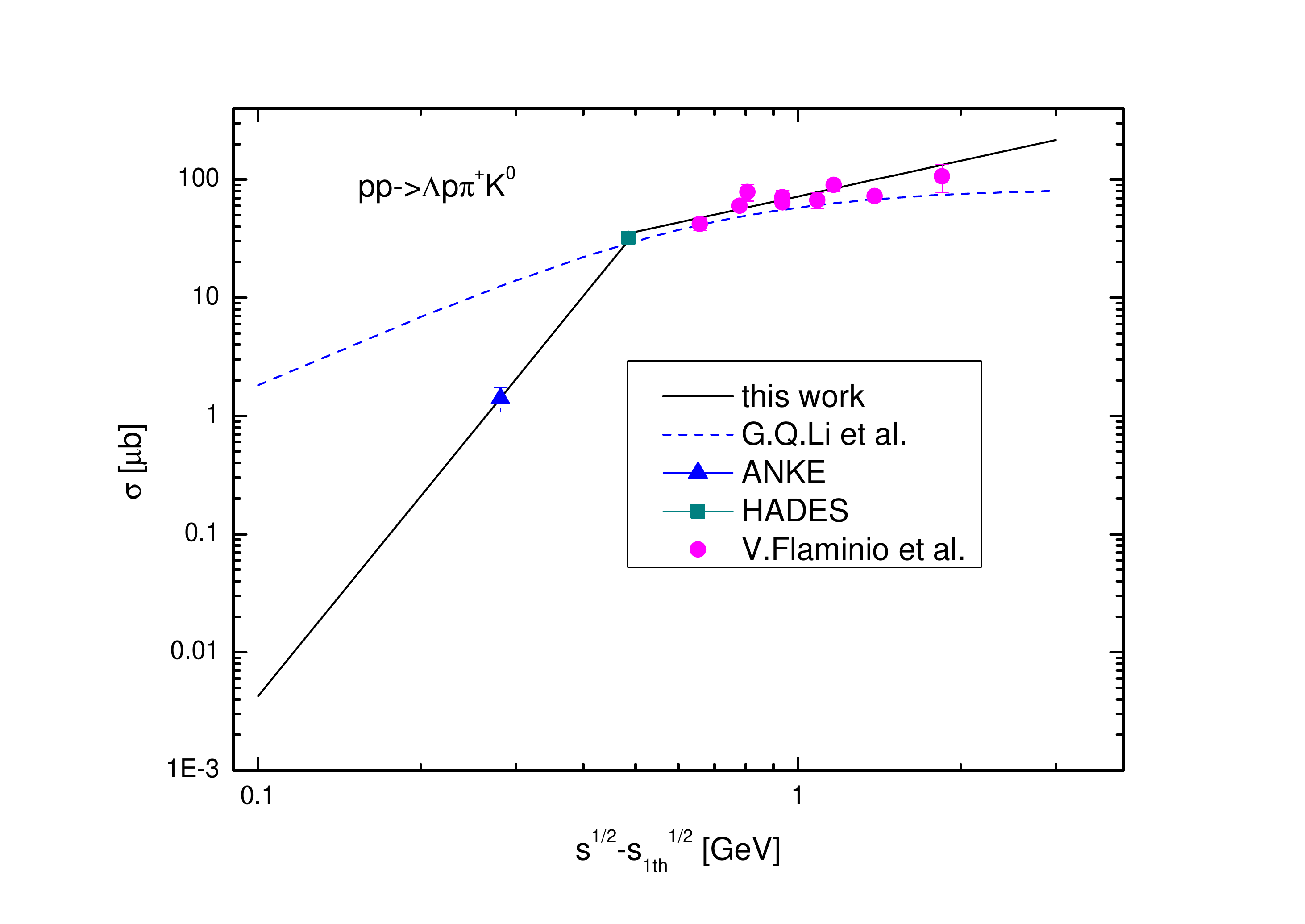}
\vspace*{-2mm} \caption{Total cross section for $pp \to {\Lambda}p{\pi^+}K^0$ reaction
as a function of excess energy. For notation see the text.}
\label{void}
\end{center}
\end{figure}
Now, we consider total cross sections of the reactions (3) with two pions in the final states.
The experimental data for these reactions are quite scarce. To date, there exist  data for the
total cross section $\sigma_{pp \to {\Lambda}p{\pi^+}{\pi^-}K^+}$
of only the process $pp \to {\Lambda}p{\pi^+}{\pi^-}K^+$ [21]. Data also are available for the total cross sections
$\sigma_{pp \to {\Lambda}p{\pi^+}{\pi^0}K^0}$ and $\sigma_{pp \to {\Lambda}n{\pi^+}{\pi^+}K^0}$ of the reactions
$pp \to {\Lambda}p{\pi^+}{\pi^0}K^0$ and $pp \to {\Lambda}n{\pi^+}{\pi^+}K^0$ respectively [21]. These data were obtained at beam energies beginning with 4.100, 4.641 and 6.045 GeV, respectively, and can be approximately fitted by the following expression suggested in Ref.~[35]:
\begin{equation}
\sigma_{pp \to {\Lambda}p{\pi^+}{\pi^-}K^+} \approx \sigma_{pp \to {\Lambda}p{\pi^+}{\pi^0}K^0} \approx
\sigma_{pp \to {\Lambda}n{\pi^+}{\pi^+}K^0} \approx
\frac{80\left(\sqrt{s}-\sqrt{s_{2{\rm th}}}\right)^{2}}{2.25+\left(\sqrt{s}-\sqrt{s_{2{\rm th}}}\right)^{2}}
~[{\rm {\mu}b}],
\end{equation}
where $\sqrt{s_{2{\rm th}}}=m_{\Lambda}+m_p+2m_{\pi^+}+m_{K^+}$ is the threshold energy.
We assume that the other two channels
$pp \to {\Lambda}p{\pi^0}{\pi^0}K^+$ and $pp \to {\Lambda}n{\pi^+}{\pi^0}K^+$ have the same cross section [35].
We will also use the parametrization (11) in the present work at incident energy of 2.83 GeV, which is lower than those studied in the compilation in Ref.~[21]
\footnote{$^)$The relations (11) are in line with the results for the
total cross sections for $K^0$ production channels $pp \to {\Lambda}p{\pi^+}{\pi^0}K^0$ and
$pp \to {\Lambda}n{\pi^+}{\pi^+}K^0$ at beam kinetic energy of 3.5 GeV, which can be obtained
assuming that these channels go exclusively through the reactions
$pp \to \Delta^{++}\Sigma(1385)^0K^0$ and $pp \to \Delta^{+}\Sigma(1385)^+K^0$
and using formula (2) for the total cross sections of the latter reactions from Ref.~[36], in which inclusive
$K^0$ production in $pp$ and $p{\rm Nb}$ interactions was measured with the HADES detector. This gives us  confidence that the use of these relations is justified at our beam energy of interest.}$^)$
.

  Finally, we focus on the total cross sections for $\Lambda$ production in $pn$ reactions (4)--(6).
Up to now, there have been no direct data on $\Lambda$ production in reaction (4).
The relationship between total cross section $\sigma_{pn \to {\Lambda}nK^+}$ of this reaction and $\sigma_{pp \to {\Lambda}pK^+}$ for channel (1) can be obtained by the following indirect route.
An analysis of the data on the production of $K^+$ mesons at small angles in proton--proton and
proton--deuteron collisions at beam energies of 1.826, 1.920, 2.020 and 2.650 GeV,
taken by the ANKE Collaboration [37], gave a value for the ratio
of total inclusive  $K^+$ creation cross sections in $pn$ and $pp$ interactions of $\sigma_{pn}^{K^+}/\sigma_{pp}^{K^+}=0.5\pm0.2$
at all four energies investigated. Since the main contribution to the total cross sections
of the reactions $pn \to K^+X$ and $pp \to K^+X$ comes, at least at initial energies $\le$ 2.020 GeV,
from the channels $pn \to {\Lambda}nK^+$ and $pp \to {\Lambda}pK^+$ [38], this means that
\begin{equation}
\sigma_{pn \to {\Lambda}nK^+} \approx \frac{1}{2} \sigma_{pp \to {\Lambda}pK^+}
\end{equation}
at these energies. Further, the $\Lambda$ hyperon production cross sections in reaction (4)
and in channel $pn \to {\Lambda}pK^0$ are the same, due to isospin symmetry.
There are [21] only five data points for the free total cross section $\sigma_{pn \to {\Lambda}pK^0}$
of this channel at incident proton kinetic energies of 5.135, 6.124 and 16.088 GeV.
Comparing these data with that available [21] for the
$pp \to {\Lambda}pK^+$ channel at similar energies (5.135, 6.045 and 11.098 GeV),
one can easily find that at these energies the $pn$ cross sections $\sigma_{pn \to {\Lambda}pK^0}$
and $\sigma_{pn \to {\Lambda}nK^+}$ are about half of the $pp$ cross sections $\sigma_{pp \to {\Lambda}pK^+}$, namely:
\begin{equation}
\sigma_{pn \to {\Lambda}pK^0}=\sigma_{pn \to {\Lambda}nK^+} \approx \frac{1}{2} \sigma_{pp \to {\Lambda}pK^+}.
\end{equation}
Accounting for expressions (12) and (13), we will assume that
\begin{equation}
\sigma_{pn \to {\Lambda}nK^+}(\sqrt{s})
\approx \frac{1}{2} \sigma_{pp \to {\Lambda}pK^+}(\sqrt{s},\sqrt{s_{{\rm th}}})
\end{equation}
also at all center-of-mass energies $\sqrt{s}$, accessible in calculation of ${\Lambda}K^+$ production
in $pA$ reactions at our beam energy of interest, 2.83 GeV, with allowance for the Fermi motion
of intranuclear nucleons
\footnote{$^)$Theoretical estimates of the ratio
$\sigma_{pn \to {\Lambda}nK^+}/\sigma_{pp \to {\Lambda}pK^+}$, obtained on the basis of
the meson-exchange models [39--41], are around 2 [39], 3 [40] or
range from 0.25 to 10 [32, 41], which is inconsistent with experimental results (12) and (13).}$^)$
.
The only set of data is available for $\Lambda$ hyperon production together with one pion in $pn$
collisions. It was obtained for $pn \to {\Lambda}p{\pi^-}K^+$ reaction at incident energies starting from 5.135 GeV [21]. The set indicates the total cross section $\sigma_{pn \to {\Lambda}p{\pi^-}K^+}$
of this reaction, which is about half of those in Eqs.~(9), (10) for $pp$ processes (2) (cf. Ref.~[35]). In line with
Ref.~[35], we will assume that the other $pn$ channel $pn \to {\Lambda}n{\pi^0}K^+$ from Eq.~(5) has
the same cross section as the $pn \to {\Lambda}p{\pi^-}K^+$ reaction. Under the above assumptions, we have:
\begin{equation}
\sigma_{pn \to {\Lambda}n{\pi^0}K^+}(\sqrt{s}) \approx \sigma_{pn \to {\Lambda}p{\pi^-}K^+}(\sqrt{s})
\approx \frac{1}{2}\sigma_{pp \to {\Lambda}p{\pi^+}K^0}({\sqrt{s},\sqrt{s_{1{\rm th}}}}).
\end{equation}
These relations (15) will be used in our calculations of $\Lambda$$K^+$ pair yields from $pA$ collisions
at all accessible values of the center-of-mass energy $\sqrt{s}$.
The only set of data there is at present for $\Lambda$ hyperon production together with two pions in $pn$
interactions. It was taken for the $pn \to {\Lambda}p{\pi^+}{\pi^-}K^0$ reaction at beam energies
of 5.135, 6.124 and 16.088 GeV [21]. The set indicates that the total cross section
$\sigma_{pn \to {\Lambda}p{\pi^+}{\pi^-}K^0}$ of this reaction is about half of those (11) for the
corresponding $pp$ channels (3) (cf. Ref.~[35]). As in Ref.~[35], we will assume that all three $pn$ processes (6)
have the same cross section as the $pn \to {\Lambda}p{\pi^+}{\pi^-}K^0$ reaction.
In this context, one gets:
\begin{equation}
\sigma_{pn \to {\Lambda}n{\pi^0}{\pi^0}K^+}(\sqrt{s})
\approx \sigma_{pn \to {\Lambda}n{\pi^+}{\pi^-}K^+}(\sqrt{s})
\approx \sigma_{pn \to {\Lambda}p{\pi^-}{\pi^0}K^+}(\sqrt{s})
\end{equation}
$$
\approx \sigma_{pn \to {\Lambda}p{\pi^+}{\pi^-}K^0}(\sqrt{s})
\approx \frac{1}{2}\sigma_{pp \to {\Lambda}p{\pi^+}{\pi^-}K^+}(\sqrt{s},\sqrt{s_{2{\rm th}}}).
$$
We will use Eq.~(16) at all collision energies of interest.
It is worth noting that, as follows from Eqs.~(7)--(16), the ratios of $\Lambda$ production
cross sections in the $pp$ and $pn$ processes (2), (3) and (5), (6) with one as well as with two pions in the final
states to those of $\Lambda$ creation channels (1) and (4) with zero outgoing pions are about of 1/9
at kinetic beam energy of 2.83 GeV of interest. This means that the main contribution to $\Lambda$
production in $pA$ reactions, even at this high incident energy, comes from the  3-body primary channels
(1) and (4).

For numerical simplicity, in the following calculations we will include
the medium modification of the final nucleon,
kaon and $\Lambda$ hyperon, participating in the production processes (1)--(6), by using
their average in-medium masses $<m^*_{h}>$ instead of their local effective masses $m^*_{h}(r)$
in the in-medium cross sections of these processes, with $<m^*_{h}>$ defined in line with Refs.~[31, 42] as:
\begin{equation}
<m^*_{h}>=m_{h}+U_h\frac{<{\rho_N}>}{{\rho_0}}.
\end{equation}
Here, $m_{h}$ is the hadron free space mass, $U_h$ is the value of its effective scalar nuclear potential at saturation
density ${\rho_0}$
\footnote{$^)$This potential is not the usual Lorentz scalar potential, but
contains both the Lorentz scalar and Lorentz vector potentials, like Eq.~(23) below.}$^)$
,
and $<{\rho_N}>$ is the average nucleon density, which was calculated separately for each
 target nucleus considered. Our calculations show that for target nuclei C, Cu, Ag and Au
the ratio $<{\rho_N}>/{\rho_0}$ is approximately equal to 0.55, 0.65, 0.72 and 0.77, respectively.
The above values will be used throughout the following study. We assume that pions do not change
their properties in the nuclear medium at densities of ordinary nuclei [43].
To match smoothly in-medium $\Lambda$ hyperon production thresholds in $pp$ collisions to those in $pn$ interactions,
we also neglect the influence of the Coulomb potentials on the final charged hadrons (protons, pions and kaons),
participating in the elementary reactions (1)--(6). In addition, these potentials, as expected, have a
minor role in the $\Lambda$ dynamics at our initial proton energy of interest.
The total energy $E^\prime_{h}$ of the hadron inside the nuclear medium can be
expressed through its average effective mass $<m^*_{h}>$ defined above and its in-medium momentum
${\bf p}^{\prime}_{h}$ as in the free particle case, namely:
\begin{equation}
E^\prime_{h}=\sqrt{(<m^*_{h}>)^2+({\bf p}^{\prime}_{h})^2}.
\end{equation}
The momentum ${\bf p}^{\prime}_{h}$ is related to the vacuum momentum ${\bf p}_{h}$
by the following expression:
\begin{equation}
E^\prime_{h}=\sqrt{(<m^*_{h}>)^2+({\bf p}^{\prime}_{h})^2}=
\sqrt{m^2_{h}+{\bf p}^2_{h}}=E_h,
\end{equation}
where $E_h$ is the hadron vacuum total energy. In the subsequent study for the $K^+$ mass shift
$U_{K^+}$ we will always employ the following option: $U_{K^+}=22$ MeV [44]. The same option will be
adopted for the $K^0$ mass shift $U_{K^0}$
\footnote{$^)$This is not true for heavy nuclei like Au where there the difference between numbers of protons and neutrons
is large due to the $\rho$ meson, which induces different mean-field potentials for $K^+$ and $K^0$
mesons when they are embedded in asymmetric nuclear matter. However, this effect is expected to be negligible
for the present approach in which the nuclear densities $\rho_N \le \rho_0$ are considered [45]. Isospin
effects lead also to only small corrections to the $K^+$ mass shift $U_{K^+}$ when going from a C to an Au
target nucleus. They are within 5\%, as our estimate, based on the $K^+p$ and $K^+n$
scattering lengths, shows.}$^)$
.
The relation between the effective scalar nucleon potential $U_N$, entering into
Eq.~(17), and the corresponding Schr${\ddot{\rm o}}$dinger equivalent potential $V_{NA}^{\rm SEP}$ (or the so-called single-particle or mean-field potential) at the normal nuclear matter density  is given by
\begin{equation}
U_N=\frac{\sqrt{m_N^2+{p^{\prime}_N}^2}}{m_N}V_{NA}^{\rm SEP}.
\end{equation}
As shown in our calculations, the vacuum momenta of the outgoing  nucleons in reactions (1)--(6) are
around $p_N$=0.6 GeV/c in the kinematics of the ANKE experiment.
Assuming that $p^{\prime}_N \approx p_N$
\footnote{$^)$This point has been numerically checked by employing Eq.~(19).}$^)$
and using $V_{NA}^{\rm SEP} \approx 0$ MeV at this in-medium nucleon momentum, corresponding to a kinetic
energy of 0.18 GeV [46],
we can readily obtain that $U_N \approx 0$ MeV. We will employ this potential throughout our present work.

   Let us now specify the effective scalar mean-field $\Lambda$ hyperon potential $U_{\Lambda}$ in Eq.~(17). A nuclear mean-field potential acting on the bound in the nucleus low-momentum
$\Lambda$ hyperon has been extracted from the properties of hypernuclei [47]. This potential has also been
 investigated in the framework of relativistic mean-field theory [48, 49]. It was found from these
studies that the well depth for a $\Lambda$ particle embedded in nuclear matter is in the vicinity of
30 MeV. The kinetic energy dependence of the $\Lambda$--nucleus mean-field potential for $\Lambda$ hyperons
colliding with a nucleus was evaluated in Ref.~[50] within the G-matrix theory over the $\Lambda$ kinetic energy
range of 0--70 MeV (or over the $\Lambda$ momentum interval of 0--0.4 GeV/c).
The momentum dependence of this potential was studied in  Ref.~[51] within the framework of the relativistic
Brueckner--Hartree--Fock theory, employing two kinds of $YN$ phenomenological potentials -- Juelich 94a and
Juelich 05 -- constructed with the meson-exchange model by the Juelich group, at the same $\Lambda$ momenta
as in Ref.~[50], and as in Ref.~[19] within the chiral unitary coupled-channel approach for $\Lambda$ momenta ranging up to 0.6 GeV/c. In both Refs.~[51] and [19], similar behavior was found for the $\Lambda$
potential at normal nuclear matter density: it is attractive and increases monotonically  with the
growth of $\Lambda$ momentum. The $\Lambda$ single-particle potential in isospin symmetric and asymmetric
nuclear matter at finite momenta up to 0.6 GeV/c has been also investigated recently in Ref.~[20], in the
framework of the Brueckner approach using the ${\Lambda}N$ potential derived from SU(3) chiral effective
field theory at next-to-leading order, as was already noted above.
Contrary to the results from Refs.~[51, 19], this potential turns from
attractive to repulsive at about 0.4 GeV/c momentum in isospin symmetric nuclear medium at saturation
density. Moreover, the $\Lambda$ single-particle potential in symmetric nuclear matter has been calculated
in Ref.~[52] within the SU$(6)$ quark model at various nuclear densities as a function of the momentum in the
range of about 0--1.4 GeV/c. An essential difference between this potential and that from Ref.~[20] is that it
turns to repulsion at fairly high momenta, around 1.1 GeV/c. Recently, the momentum and density dependences
of the $\Lambda$ hyperon single-particle potential in nuclear medium was calculated starting from QCD
on the lattice [53]. The reported results are compatible with those obtained within other approaches.
Since the accessible range of the $\Lambda$
hyperon momenta in the ANKE experiment is about 0.4--3 GeV/c, it is helpful to also estimate the $\Lambda$
mean-field potential, needed for our present study, for such high $\Lambda$ momenta.
We will rely on the constituent quark model, which has been employed in Ref.~[54] to derive the density dependence
of the low-momentum $\Lambda$--nucleus potential, and will proceed analogously with the aim of obtaining its
momentum dependence at saturation density $\rho_0$. In this model, the $\Lambda$ mean-field scalar
$U_{S{\Lambda}}$ and vector $U_{V{\Lambda}}$ potentials are about 2/3 of
$U_{SN}$ and $U_{VN}$ of a nucleon when in-medium nucleon and $\Lambda$ hyperon velocities
$v^{\prime}_N$ and $v^{\prime}_{\Lambda}$ relative to the nuclear matter are equal to each other, i.e.,
\begin{equation}
U_{S{\Lambda}}(v^{\prime}_{\Lambda},\rho_N)=\frac{2}{3}U_{SN}(v^{\prime}_{N},\rho_N),\,\,\,
U_{V{\Lambda}}(v^{\prime}_{\Lambda},\rho_N)=\frac{2}{3}U_{VN}(v^{\prime}_{N},\rho_N);\,\,\,
v^{\prime}_{N}=v^{\prime}_{\Lambda}.
\end{equation}
The latter term in Eq.~(21) corresponds, as is easy to see, to the following relation between
the respective in-medium nucleon momentum $p^{\prime}_N$ and the $\Lambda$ momentum $p^{\prime}_{\Lambda}$:
\begin{equation}
p^{\prime}_{N}=\frac{<m^*_{N}>}{<m^*_{\Lambda}>}p^{\prime}_{\Lambda}.
\end{equation}
However, for reasons of numerical simplicity, calculating the $\Lambda$--nucleus mean-field potential, we will
use the free space nucleon and $\Lambda$ hyperon masses $m_N$ and $m_{\Lambda}$ in Eq.~(22)  instead of
their average in-medium masses $<m^*_{N}>$ and $<m^*_{\Lambda}>$.
Then, this potential $V_{{\Lambda}A}^{\rm SEP}$ can be
defined as [54]
\footnote{$^)$The space-like component of the $\Lambda$ vector self-energy is ignored here.}$^)$
:
\begin{equation}
V_{{\Lambda}A}^{\rm SEP}(p^{\prime}_{\Lambda},\rho_N)=
\sqrt{\left[m_{\Lambda}+U_{S{\Lambda}}(p^{\prime}_{\Lambda},\rho_N)\right]^2+{p^{\prime}_{\Lambda}}^2}
+U_{V{\Lambda}}(p^{\prime}_{\Lambda},\rho_N)-\sqrt{m_{\Lambda}^2+{p^{\prime}_{\Lambda}}^2}.
\end{equation}
Adopting the momentum-dependent parametrization for the nucleon scalar and vector potentials
at saturation density $\rho_0$ from Ref.~[46],
\begin{equation}
U_{SN}(p^{\prime}_{N},\rho_0)=-\frac{494.2272}{1+0.3426\sqrt{p^{\prime}_{N}/p_F}}~{\rm MeV},
\end{equation}
\begin{equation}
U_{VN}(p^{\prime}_{N},\rho_0)=\frac{420.5226}{1+0.4585\sqrt{p^{\prime}_{N}/p_F}}~{\rm MeV}
\end{equation}
(where $p_F=1.35$ fm$^{-1}=$0.2673 GeV/c)
and using Eqs.~(21)--(23), we calculated the momentum dependence of potential
$V_{{\Lambda}A}^{\rm SEP}$ at density $\rho_0$
\footnote{$^)$Adopting the momentum-dependent parametrization for the nucleon scalar and vector potentials
at different nuclear densities from Ref.~[46] and using Eq.~(23), it is easy to also calculate the
density dependence of potential $V_{{\Lambda}A}^{\rm SEP}$ at different lambda momenta. However, this is beyond
the scope of the present work, since here our attention is focussed mainly on the $\Lambda$ in-medium
properties at saturation density $\rho_0$.}$^)$
.
It is shown  by the dashed
curve in Fig. 3. We have also made an adjustment by multiplying the vector $\Lambda$ hyperon potential by a factor
of 1.068, to get a value for the potential $V_{{\Lambda}A}^{\rm SEP}$ at zero momentum consistent with the
experimental value of -(32$\pm$2) MeV (full circle in figure 3), extracted from the data on binding energies
of $\Lambda$ single-particle states in nuclei [47]. The potential adjusted in this way
is presented by the solid curve in Fig. 3. In this case the $\Lambda$--nucleus potential
is attractive for momenta $\le$ 0.7 GeV/c, whereas it becomes repulsive for higher momenta and reaches
the value $\approx$ 70 MeV at $\Lambda$ momentum of 3 GeV/c.
\begin{figure}[!ht]
\begin{center}
\includegraphics[width=12.0cm]{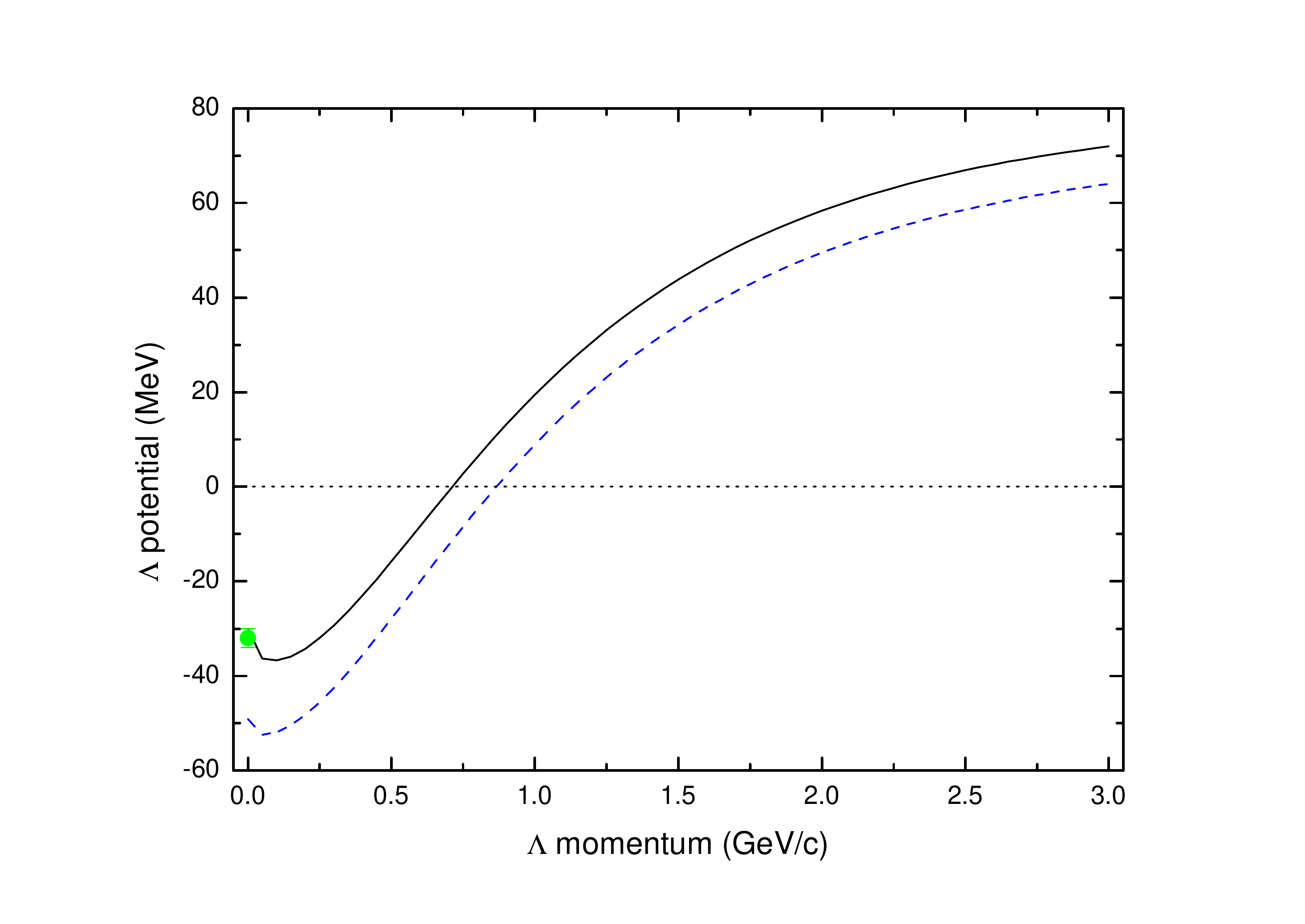}
\vspace*{-2mm} \caption{Schr$\ddot{\rm o}$dinger equivalent ${\Lambda}$ hyperon potential at density $\rho_0$ as a function of $\Lambda$ momentum relative to the nuclear matter at rest calculated on the
basis of Eq.~(23), without and with rescaling its vector potential by a factor of 1.068
(dashed and solid lines, respectively).}
\label{void}
\end{center}
\end{figure}

As follows from Figs. 5--10,
the coincident ${\Lambda}K^+$ yield is appreciably sensitive to the
lambda potential at vacuum momenta around $p_{\Lambda}=1$ GeV/c,
where the secondary pion--nucleon ${\Lambda}K^+$ production channels (49) and (50) considered below
are dominant. Assuming that, as in the nucleon case above, $p^{\prime}_{\Lambda} \approx p_{\Lambda}$ and using
the results shown in Fig. 3, we find that $V_{{\Lambda}A}^{\rm SEP} \approx 20$ MeV for this in-medium lambda momentum. Then, taking into account that the relation between the effective scalar hyperon potential $U_{\Lambda}$,
in Eq.~(17), and the corresponding Schr${\ddot{\rm o}}$dinger equivalent potential
$V_{{\Lambda}A}^{\rm SEP}$ at normal nuclear matter density  is given by the relation analogous to Eq.~(20)
for nucleons, we can readily obtain that $U_{\Lambda} \approx 30$ MeV at the above $\Lambda$ momentum.
Since the ${\Lambda}K^+$ yield from the  direct ${\Lambda}K^+$
production mechanisms considered is concentrated mainly at $\Lambda$ momenta around
$p_{\Lambda}$ $\approx$ 2.6 GeV/c, for which it reacts weakly on the $\Lambda$--nuclear potential, as follows from Figs. 7--10,
we will adopt the value $U_{\Lambda} \approx 30$ MeV in our calculations of this yield.
However, in reality it is still unclear which $\Lambda$--nucleus potential is the correct one for
such high $\Lambda$ momenta (cf., for instance, results from Refs. [20] and [52]).
Therefore, to extend the range of applicability of our model, we will also calculate the
$\Lambda$ production cross sections off nuclei  in  scenarios with  possible $\Lambda$ mass shift
(or effective scalar potential) $U_{\Lambda}$ ranging from -30 MeV to 60 MeV.

  Accounting for the distortions of the initial proton and final $\Lambda$ hyperon and kaon as well as
the fact that in the ANKE experiment the latter were detected in the forward polar angular domains
$0^{\circ} \le \theta_{\Lambda} \le 6^{\circ}$ and $0^{\circ} \le \theta_{K^+} \le 12^{\circ}$, respectively, and using the results given in Refs.~[55--58], we can represent
the differential cross section for the production on nuclei of $\Lambda$ hyperons with the vacuum momentum
${\bf p}_{\Lambda}$ in coincidence with $K^+$ mesons with the vacuum momentum ${\bf p}_{K^+}$
in the primary proton-induced reaction channels (1)--(6) as follows:
\begin{equation}
\frac{d\sigma_{pA\to {\Lambda}K^+X}^{({\rm prim})}
({\bf p}_0,{\bf p}_{\Lambda},{\bf p}_{K^+})}
{d{\bf p}_{\Lambda}d{\bf p}_{K^+}}=I_{K^+{\Lambda}}[A]
\end{equation}
$$
\times
\left[\frac{Z}{A}
\left<\frac{d\sigma_{pp\to {\Lambda}K^+X}({\bf p}^{\prime}_{0},{\bf p}^{\prime}_{\Lambda},
{\bf p}^{\prime}_{K^+})}{d{\bf p}^{\prime}_{\Lambda}d{\bf p}^{\prime}_{K^+}}\right>_A+
\frac{N}{A}
\left<\frac{d\sigma_{pn\to {\Lambda}K^+X}({\bf p}^{\prime}_{0},
{\bf p}^{\prime}_{\Lambda},{\bf p}^{\prime}_{K^+})}{d{\bf p}^{\prime}_{\Lambda}d{\bf p}^{\prime}_{K^+}}\right>_A
\right]\frac{d{\bf p}^{\prime}_{\Lambda}}{d{\bf p}_{\Lambda}}\frac{d{\bf p}^{\prime}_{K^+}}{d{\bf p}_{K^+}},
$$
where
\begin{equation}
I_{K^+{\Lambda}}[A]=2{\pi}A\int\limits_{0}^{R}r_{\bot}dr_{\bot}
\int\limits_{-\sqrt{R^2-r_{\bot}^2}}^{\sqrt{R^2-r_{\bot}^2}}dz
\rho(\sqrt{r_{\bot}^2+z^2})
\end{equation}
$$
\times
\exp{\left[-\sigma_{pN}^{\rm in}A\int\limits_{-\sqrt{R^2-r_{\bot}^2}}^{z}\rho(\sqrt{r_{\bot}^2+x^2})dx
-\sigma_{K^+N}^{\rm tot}A\int\limits_{z}^{\sqrt{R^2-r_{\bot}^2}}
\rho(\sqrt{r_{\bot}^2+x^2})dx\right]}
$$
$$
\times
\exp{\left[-\sigma_{{\Lambda}N}^{\rm tot}(p^{\prime}_{\Lambda})A
\int\limits_{z}^{\sqrt{R^2-r_{\bot}^2}}
\rho(\sqrt{r_{\bot}^2+x^2})dx\right]},
$$
\begin{equation}
\left<\frac{d\sigma_{pN\to {\Lambda}K^+X}({\bf p}^{\prime}_{0},{\bf p}^{\prime}_{\Lambda},{\bf p}^{\prime}_{K^+})}
{d{\bf p}^{\prime}_{\Lambda}d{\bf p}^{\prime}_{K^+}}\right>_A
=
\int\int
P_A({\bf p}_t,E)d{\bf p}_tdE
\end{equation}
$$
\times
\left\{\frac{d\sigma_{pN\to {\Lambda}K^+X}[\sqrt{s},<m^*_{K^+}>,<m^*_{N}>,<m^*_{\Lambda}>,{\bf p}^{\prime}_{\Lambda},
{\bf p}^{\prime}_{K^+}]}{d{\bf p}^{\prime}_{\Lambda}d{\bf p}^{\prime}_{K^+}}\right\}
$$
and
\begin{equation}
p^{\prime}_{\Lambda}=\sqrt{E^2_{\Lambda}-(<m^*_{\Lambda}>)^2},\,\,\,
E_{\Lambda}=\sqrt{m_{\Lambda}^2+{\bf p}_{\Lambda}^2}.
\end{equation}
Here,
$d\sigma_{pN\to {\Lambda}K^+X}[\sqrt{s},<m^*_{K^+}>,<m^*_{N}>,<m^*_{\Lambda}>,
{\bf p}^{\prime}_{\Lambda},{\bf p}^{\prime}_{K^+}]/d{\bf p}^{\prime}_{\Lambda}d{\bf p}^{\prime}_{K^+}$
are the ``in-medium" differential cross sections for the production of $\Lambda$ hyperons and $K^+$ mesons
with the in-medium momenta ${\bf p}^{\prime}_{\Lambda}$ and ${\bf p}^{\prime}_{K^+}$, correspondingly, in reactions
(1)--(3) ($N=p$) and in (4)--(6) ($N=n$) at the $pN$ center-of-mass energy $\sqrt{s}$;
$\rho(r)$ and $P_A({\bf p}_t,E)$ are the local nucleon density and the
spectral function of the target nucleus $A$ normalized to unity, respectively;
${\bf p}_{t}$  and $E$ are the internal momentum and binding energy of the struck target nucleon
just before the collision, respectively; $\sigma_{pN}^{\rm in}$ and
$\sigma_{{K^+}N}^{\rm tot}$, $\sigma_{{\Lambda}N}^{\rm tot}$
are the inelastic and total cross sections of the $pN$ and $K^+N$, ${\Lambda}N$ interactions respectively;
$Z$ and $N$ are the numbers of protons and neutrons respectively in the target nucleus ($A=Z+N$), and $R$ is its radius; and
${\bf p}_{0}$ and ${\bf p}^{\prime}_{0}$ are the momenta of the initial proton outside and inside
the target nucleus respectively. They are linked by the equation [57]:
\begin{equation}
 {\bf p}_{0}^{'}={\bf p}_{0}-\Delta{\bf p},\,\,\,
 \Delta{\bf p}=\frac{E_0V_0}{p_0}\frac{{\bf p}_{0}}{|{\bf p}_{0}|},
\end{equation}
where $E_0$ and $V_0$ are the total energy of the initial proton outside the nucleus
and the nuclear optical potential that this proton feels in the interior of the nucleus
($V_0 \approx 40~{\rm MeV}$) respectively. The expression for $s$ is given by the formula [57]:
\begin{equation}
  s=(E_{0}^{'}+E_t)^2-({\bf p}_{0}^{'}+{\bf p}_t)^2,
\end{equation}
where
\begin{equation}
 E_{0}^{'}=E_{0}-\frac{{\Delta{\bf p}}^2}{2M_A},
\end{equation}
\begin{equation}
   E_t=M_A-\sqrt{(-{\bf p}_t)^2+(M_{A}-m_{N}+E)^{2}}.
\end{equation}
Here, $M_A$ is the mass of the target nucleus.

For the $\Lambda$$K^+$ production calculations in the cases of the $^{12}$C, $^{63}$Cu, $^{108}$Ag, and $^{197}$Au
target nuclei reported here, we have employed for the nuclear density $\rho({\bf r})$,
respectively, the harmonic oscillator and the Woods-Saxon distributions:
\begin{equation}
\rho({\bf r})={\rho}_{N}({\bf r})/A=\frac{(b/\pi)^{3/2}}{A/4}\left\{1+
\left[\frac{A-4}{6}\right]br^{2}\right\}\exp{(-br^2)},
\end{equation}
\begin{equation}
 \rho({\bf r})=\rho_{0}\left[1+
\exp{\left(\frac{r-R_{1/2}}{a}\right)}\right]^{-1}
\end{equation}
with $b=0.355~{\rm fm}^{-2}$ and $R_{1/2}=4.20~{\rm fm}$ for $^{63}$Cu,
$R_{1/2}=5.505~{\rm fm}$ for $^{108}$Ag, $R_{1/2}=6.825~{\rm fm}$ for $^{197}$Au,
and $a=0.55~{\rm fm}$ for all nuclei [55, 59].
The nuclear spectral function $P_A({\bf p}_t,E)$ (which represents the
probability of finding a nucleon with momentum ${\bf p}_t$ and removal energy $E$ in the nucleus) for the $^{12}$C target nucleus was taken from Ref.~[60].
The single-particle part of this function for the $^{63}$Cu, $^{108}$Ag and $^{197}$Au
target nuclei was assumed to be the same as that for $^{208}$Pb [44].
The latter was taken from Ref.~[61]. The correlated part of the
nuclear spectral function for these target nuclei was borrowed from Ref.~[60].
Using the total cross sections $\sigma_{{K^+}N}^{\rm tot}$ and $\sigma_{{\Lambda}N}^{\rm tot}$ in Eq.~(27), we assume that if a kaon or a $\Lambda$ hyperon undergoes a quasi-elastic collision with the target nucleon, it will not fall in the ANKE acceptance window.

    Taking into account the above arguments that the main contribution to
the primary ${\Lambda}K^+$ production in $pA$ collisions at the considered beam energy of 2.83 GeV will come from the three-body direct processes (1) and (4), we describe the in-medium differential cross sections
$d\sigma_{pN\to {\Lambda}K^+X}[\sqrt{s},<m^*_{K^+}>,<m^*_{N}>,<m^*_{\Lambda}>,
{\bf p}^{\prime}_{\Lambda},{\bf p}^{\prime}_{K^+}]/d{\bf p}^{\prime}_{\Lambda}d{\bf p}^{\prime}_{K^+}$
according to the three-body phase space. Following Ref.~[57], we obtain:
\begin{equation}
\frac{d\sigma_{pN\to {\Lambda}K^+X}[\sqrt{s},<m^*_{K^+}>,<m^*_{N}>,<m^*_{\Lambda}>,
{\bf p}^{\prime}_{\Lambda},{\bf p}^{\prime}_{K^+}]}{d{\bf p}^{\prime}_{\Lambda}d{\bf p}^{\prime}_{K^+}}=
\frac{1}{8E^{\prime}_{\Lambda}E^{\prime}_{K^+}}
\end{equation}
$$
\times
\frac{\sigma_{pN\to {\Lambda}K^+X}(\sqrt{s},\sqrt{s^*_{\rm th}},
\sqrt{s^*_{1{\rm th}}},\sqrt{s^*_{2{\rm th}}})}{I_3(s,<m^*_{K^+}>,<m^*_N>,<m^*_{\Lambda}>)}
\frac{1}{(\omega+E_t)}
\delta\left(\omega+E_t-
\sqrt{(<m^*_{N}>)^2+({\bf Q}+{\bf p}_t)^2}\right),
$$
where
\begin{equation}
\omega=E^{\prime}_{0}-E^{\prime}_{\Lambda}-E^{\prime}_{K^+},\,\,\,
{\bf Q}={\bf p}^{\prime}_{0}-{\bf p}^{\prime}_{\Lambda}-{\bf p}^{\prime}_{K^+}
\end{equation}
and the quantity $I_3$ is defined as [57]:
\begin{equation}
I_{3}(s,<m^*_{K^+}>,<m^*_{N}>,<m^*_{\Lambda}>)=(\frac{{\pi}}{2})^2
\int\limits_{(<m^*_{K^+}>+<m^*_{N}>)^2}^{({\sqrt{s}}-<m^*_{\Lambda}>)^2}
\frac{\lambda[x,(<m^*_{K^+}>)^{2},(<m^*_{N}>)^{2}]}{x}
\end{equation}
$$
\times
\frac{\lambda[s,x,(<m^*_{\Lambda}>)^{2}]}{s}\,dx,
$$
\begin{equation}
\lambda[x,y,z]=\sqrt{\left[x-({\sqrt{y}}+{\sqrt{z}})^2\right]\left[x-
({\sqrt{y}}-{\sqrt{z}})^2\right]}.
\end{equation}
Here,
$\sigma_{pN\to {\Lambda}K^+X}(\sqrt{s},\sqrt{s^*_{\rm th}},\sqrt{s^*_{1{\rm th}}},\sqrt{s^*_{2{\rm th}}})$
are the ``in-medium" total cross sections for the ${\Lambda}K^+$ production in reactions (1)--(3)
($N=p$) as well as in (4)--(6) ($N=n$), with the threshold energies
$\sqrt{s^*_{\rm th}}=<m^*_{\Lambda}>+<m^*_{p}>+<m^*_{K^+}>$,
$\sqrt{s^*_{1{\rm th}}}=<m^*_{\Lambda}>+<m^*_{p}>+m_{\pi^+}+<m^*_{K^0}>$
and $\sqrt{s^*_{2{\rm th}}}=<m^*_{\Lambda}>+<m^*_{p}>+2m_{\pi^+}+<m^*_{K^+}>$.
As in Ref.~[58], we assume that these cross sections are equivalent to the vacuum cross sections
$\sigma_{pN\to {\Lambda}K^+X}(\sqrt{s},\sqrt{s_{\rm th}},\sqrt{s_{1{\rm th}}},\sqrt{s_{2{\rm th}}})$
in which the free threshold energies $\sqrt{s_{\rm th}}$, $\sqrt{s_{1{\rm th}}}$ and $\sqrt{s_{2{\rm th}}}$
are replaced by the in-medium threshold energies $\sqrt{s^*_{\rm th}}$, $\sqrt{s^*_{1{\rm th}}}$ and
$\sqrt{s^*_{2{\rm th}}}$. Due to the above considerations, the vacuum cross sections
$\sigma_{pp\to {\Lambda}K^+X}(\sqrt{s},\sqrt{s_{\rm th}},\sqrt{s_{1{\rm th}}},\sqrt{s_{2{\rm th}}})$
and
$\sigma_{pn\to {\Lambda}K^+X}(\sqrt{s},\sqrt{s_{\rm th}},\sqrt{s_{1{\rm th}}},\sqrt{s_{2{\rm th}}})$
can be defined as:
\begin{equation}
\sigma_{pp\to {\Lambda}K^+X}(\sqrt{s},\sqrt{s_{\rm th}},\sqrt{s_{1{\rm th}}},\sqrt{s_{2{\rm th}}})=
\sigma_{pp\to {\Lambda}pK^+}(\sqrt{s},\sqrt{s_{\rm th}})+
2\sigma_{pp\to {\Lambda}p{\pi^+}K^0}(\sqrt{s},\sqrt{s_{1{\rm th}}})
\end{equation}
$$
+
3\sigma_{pp\to {\Lambda}p{\pi^+}{\pi^-}K^+}(\sqrt{s},\sqrt{s_{2{\rm th}}}),
$$
\begin{equation}
\sigma_{pn\to {\Lambda}K^+X}(\sqrt{s},\sqrt{s_{\rm th}},\sqrt{s_{1{\rm th}}},\sqrt{s_{2{\rm th}}})=
\frac{1}{2}\sigma_{pp\to {\Lambda}pK^+}(\sqrt{s},\sqrt{s_{\rm th}})+
\sigma_{pp\to {\Lambda}p{\pi^+}K^0}(\sqrt{s},\sqrt{s_{1{\rm th}}})
\end{equation}
$$
+
\frac{3}{2}\sigma_{pp\to {\Lambda}p{\pi^+}{\pi^-}K^+}(\sqrt{s},\sqrt{s_{2{\rm th}}})=
\frac{1}{2}\sigma_{pp\to {\Lambda}K^+X}(\sqrt{s},\sqrt{s_{\rm th}},\sqrt{s_{1{\rm th}}},\sqrt{s_{2{\rm th}}}).
$$

We now specify the cross sections $\sigma_{pN}^{\rm in}$, $\sigma_{{K^+}N}^{\rm tot}$
and $\sigma_{{\Lambda}N}^{\rm tot}$, in Eq.~(27). We use $\sigma_{pN}^{\rm in}=30$ mb for the incident proton energy and $\sigma_{K^+N}^{\rm tot}=12$ mb for all kaon momenta
involved in our calculations [58].
Due to the isospin symmetry, the total cross sections $\sigma_{{\Lambda}p}^{\rm tot}$ and
$\sigma_{{\Lambda}n}^{\rm tot}$ of the free ${\Lambda}p$ and ${\Lambda}n$ interactions are the same
and we denote them as $\sigma_{{\Lambda}N}^{\rm tot}$. At $\Lambda$ momenta of a few GeV/c of interest
the cross section $\sigma_{{\Lambda}p}^{\rm tot}$ is entirely exhausted, as  our calculations show, by
the total cross sections $\sigma_{{\Lambda}p \to {\Lambda}p}$ and $\sigma_{{\Lambda}p \to {\Sigma}^0p}$,
$\sigma_{{\Lambda}p \to {\Sigma}^+n}$ of elastic ${\Lambda}p \to {\Lambda}p$ and inelastic
${\Lambda}p \to {\Sigma}^0p$, ${\Lambda}p \to {\Sigma}^+n$ processes. The isospin considerations show that
$\sigma_{{\Lambda}p \to {\Sigma}^+n}=2\sigma_{{\Lambda}p \to {\Sigma}^0p}$. With these, we have:
\begin{equation}
\sigma_{{\Lambda}N}^{\rm tot}=\sigma_{{\Lambda}p}^{\rm tot}=\sigma_{{\Lambda}p \to {\Lambda}p}
+3\sigma_{{\Lambda}p \to {\Sigma}^0p}.
\end{equation}
For the free lambda-proton total cross sections $\sigma_{{\Lambda}p \to {\Lambda}p}$ and
$\sigma_{{\Lambda}p \to {\Sigma}^0p}$ as functions of the laboratory $\Lambda$ momentum $p_{\Lambda}$,
we employ the following parametrizations suggested in Ref.~[62]:
\begin{equation}
\sigma_{{\Lambda}p \to {\Lambda}p}(p_{\Lambda})=(39.66-100.45x+92.44x^2-21.40x^3)/p_{\Lambda}~[{\rm mb}],
\end{equation}
\begin{equation}
\sigma_{{\Lambda}p \to {\Sigma}^0p}(p_{\Lambda})=(31.10-30.94x+8.16x^2)
(p_{\Sigma}^{\rm cm}/p_{\Lambda}^{\rm cm})~[{\rm mb}],
\end{equation}
where $x=Min(2.1~{\rm GeV/c}$, and $p_{\Lambda})$, $p_{\Lambda}^{\rm cm}$ and $p_{\Sigma}^{\rm cm}$ are the corresponding cm momenta. In Eqs.~(43) and (44), the momenta are expressed in GeV/c.
For the in-medium ${\Lambda}p$ total cross section we use Eqs.~(42)--(44), in which cm momenta
$p_{\Lambda}^{\rm cm}$ and $p_{\Sigma}^{\rm cm}$ are defined as follows:
\begin{equation}
p_{\Lambda}^{\rm cm}=\frac{1}{2\sqrt{s_{\Lambda}}}\lambda[s_{\Lambda},m_N^2,(<m_{\Lambda}^*>)^2],\,\,
p_{\Sigma}^{\rm cm}=\frac{1}{2\sqrt{s_{\Sigma}}}\lambda[s_{\Sigma},m_N^2,m_{\Sigma^0}^2],\,\,
s_{\Sigma}=s_{\Lambda}=(E^{\prime}_{\Lambda}+m_N)^2-{p^{\prime}_{\Lambda}}^2,
\end{equation}
where $m_{\Sigma^0}$ is the $\Sigma^0$ hyperon free space mass
\footnote{$^)$Following the predictions of the chiral effective field theory approach [20, 63] and
SU$(6)$ quark model [52] for the fate of hyperons in nuclear matter and phenomenological information
inferred from hypernuclear data [1, 64] that the $\Sigma$ hyperon experiences only a moderately repulsive nuclear potential of order 10--40 MeV at central nuclear densities and finite momenta as well as a
weakly attractive one at the surface of the nucleus, we assume that the mass of the $\Sigma^0$ hyperon
is not changed in the nuclear medium.}$^)$
and the quantity $\lambda[x,y,z]$ is defined above by Eq.~(39).

  Let us now modify  Eq.~(26), describing the respective differential cross section for
${\Lambda}K^+$ production in $pA$ collisions from primary processes (1)--(6), to that corresponding to the kinematical conditions of the ANKE experiment. In this experiment, the differential cross section
for production of $\Lambda$ hyperons in the polar angular range of $0^{\circ} \le \theta_{\Lambda} \le 6^{\circ}$
in the lab system in the interaction of protons of energy of 2.83 GeV with the C, Cu, Ag, and Au target nuclei
in coincidence with $K^+$ mesons, which were required to have vacuum momenta in the interval of
0.2 GeV/c $\le p_{K^+} \le$ 0.6 GeV/c and to be in the polar angular domain of
$0^{\circ} \le \theta_{K^+} \le 12^{\circ}$, was measured as a function of their vacuum momentum.
Performing the respective integration of the full differential cross section (26) over the ANKE acceptance window,
we can represent this differential cross section in the following form:
\begin{equation}
\left<\frac{d\sigma_{pA\to {\Lambda}X}^{({\rm prim})}
({\bf p}_0,p_{\Lambda})}
{dp_{\Lambda}d{\bf \Omega}_{\Lambda}}\right>_
{\Delta{\bf \Omega}_{\Lambda}\Delta{\bf p}_{K^+}}=
\frac{1}{(2\pi)(1-\cos{6^{\circ}})}
\end{equation}
$$
\times
\int\limits_{0.2~{\rm GeV/c}}^{0.6~{\rm GeV/c}}dp_{K^+}
\int\limits_{\cos{12^{\circ}}}^{1}d\cos{\theta_{K^+}}
\int\limits_{\cos{6^{\circ}}}^{1}d\cos{\theta_{\Lambda}}
\int\limits_{0}^{2\pi}d\phi_{K^+}
\int\limits_{0}^{2\pi}d\phi_{\Lambda}
$$
$$
\times
\frac{d\sigma_{pA\to {\Lambda}K^+X}^{({\rm prim})}
({\bf p}_0,{\bf p}_{\Lambda},{\bf p}_{K^+})}
{d{\bf p}_{\Lambda}d{\bf p}_{K^+}}p^2_{\Lambda}p^2_{K^+},
$$
where
\begin{equation}
\Delta{\bf \Omega}_{\Lambda}=2\pi(1-\cos{6^{\circ}}),\,\,\,
\Delta{\bf p}_{K^+}: 0.2~{\rm GeV/c} \le p_{K^+} \le 0.6~{\rm GeV/c},\\\
0^{\circ} \le \theta_{K^{+}} \le 12^{\circ}.\\\
\end{equation}
Here, $\phi_{K^+}$ and $\phi_{\Lambda}$ are the azimuthal angles of the kaon and $\Lambda$ hyperon
momenta ${\bf p}_{K^+}$ and ${\bf p}_{\Lambda}$ in the lab system.

\section*{2.2 Two-step ${\Lambda}K^+$ production mechanisms}

\hspace{1.5cm} At our incident energy of  interest, the following two-step processes
with pions in an intermediate states contribute mainly (see below) to the ${\Lambda}K^+$ production in $pA$ reactions
\footnote{$^)$We assume that the $\Delta$ resonance, produced by first-chance $pN$ collisions, decays into $\pi$
and nucleon immediately after its production in these collisions. This assumption is well justified due to the
following. The $\Delta$ decay mean free path can be evaluated as
$\lambda_{\Delta}^{\rm dec}=p_{\Delta}/(m_{\Delta}{\Gamma_{\Delta}})$, where $p_{\Delta}$, $m_{\Delta}$ and
$\Gamma_{\Delta}$ are the $\Delta$ resonance laboratory momentum, pole mass and width, respectively.
For typical values $p_{\Delta} \approx m_{\Delta}$ and $\Gamma_{\Delta} \approx 120$ MeV, we obtain that
$\lambda_{\Delta}^{\rm dec} \approx 1.7$ fm. The $\Delta$ mean free path up to inelastic interaction
can be estimated as
$\lambda_{\Delta}^{\rm in}=1/(<\rho_N>\sigma_{{\Delta}N}^{\rm in})$, where $\sigma_{{\Delta}N}^{\rm {in}}$
is the appropriate ${\Delta}N$ inelastic cross section. Using $<\rho_N>=0.55 \rho_0$ ($^{12}$C target
nucleus), $\rho_0=0.16$ fm$^{-3}$, and as an estimate of $\sigma_{{\Delta}N}^{\rm {in}}$ the relation
$\sigma_{{\Delta}N}^{\rm {in}}=\frac{3}{4}\sigma_{pN}^{\rm {in}}$ (cf. second Ref. from Ref.~[11]) with
value $\sigma_{pN}^{\rm {in}}=30$ mb, we get that $\lambda_{\Delta}^{\rm in} \approx 5.1$ fm.
The latter value is three times larger than the estimated above $\Delta$ decay mean free path.}$^)$
:
\begin{equation}
p+N \to \pi^+,\pi^0,\pi^-+X;
\end{equation}
\begin{eqnarray}
\pi^++n \to \Lambda+K^+,\nonumber\\
\pi^0+p \to \Lambda+K^+;
\end{eqnarray}
\begin{eqnarray}
\pi^++p \to \Lambda+\pi^++K^+,\nonumber\\
\pi^0+p \to \Lambda+\pi^0+K^+,\nonumber\\
\pi^-+p \to \Lambda+\pi^-+K^+,\nonumber\\
\pi^++n \to \Lambda+\pi^0+K^+,\nonumber\\
\pi^0+n \to \Lambda+\pi^-+K^+.
\end{eqnarray}
Remember that the free threshold energies (or momenta), e.g., for the processes
${\pi^+}n \to {\Lambda}K^+$ and ${\pi^+}n \to {\Lambda}{\pi^0}K^+$,
respectively, are 0.76 (0.89) and 1.0 GeV (1.13 GeV/c).

Adopting the results given in Refs.~[58, 59], the
differential ${\Lambda}K^+$ production cross section for $pA$ collisions at small laboratory angles
from the secondary channels (49) and (50) can be represented as follows:
\begin{equation}
\frac{d\sigma_{pA\to {\Lambda}K^+X}^{({\rm sec}),(\pi)}
({\bf p}_0,{\bf p}_{\Lambda},{\bf p}_{K^+})}
{d{\bf p}_{\Lambda}d{\bf p}_{K^+}}=\frac{I^{\rm sec}_{V}[A]}{I^{\prime}_{V}[A]}
\sum_{\pi^{\prime}=\pi^+,\pi^0,\pi^-}\int \limits_{4\pi}d{\bf \Omega}_{\pi}
\int \limits_{p_{\pi}^{{\rm abs}}}^{p_{\pi}^{{\rm lim}}
(\vartheta_{\pi})}p_{\pi}^{2}
dp_{\pi}
\frac{d\sigma_{pA\to {\pi^{\prime}}X}^{({\rm prim})}({\bf p}_0)}{d{\bf p}_{\pi}}
\end{equation}
$$
\times
\left[\frac{Z}{A}\left<\frac{d\sigma_{{\pi^{\prime}}p \to {\Lambda}K^+X}({\bf p}_{\pi},
{\bf p}^{\prime}_{\Lambda},{\bf p}^{\prime}_{K^+})}{d{\bf p}^{\prime}_{\Lambda}d{\bf p}^{\prime}_{K^+}}\right>_A+
\frac{N}{A}\left<\frac{d\sigma_{{\pi^{\prime}}n \to {\Lambda}K^+X}({\bf p}_{\pi},
{\bf p}^{\prime}_{\Lambda},{\bf p}^{\prime}_{K^+})}{d{\bf p}^{\prime}_{\Lambda}
d{\bf p}^{\prime}_{K^+}}\right>_A\right]
\frac{d{\bf p}^{\prime}_{\Lambda}}{d{\bf p}_{\Lambda}}
\frac{d{\bf p}^{\prime}_{K^+}}{d{\bf p}_{K^+}},
$$
where
\begin{equation}
I^{\rm sec}_{V}[A]=2{\pi}A^2\int\limits_{0}^{R}r_{\bot}dr_{\bot}
\int\limits_{-\sqrt{R^2-r_{\bot}^2}}^{\sqrt{R^2-r_{\bot}^2}}dz
\rho(\sqrt{r_{\bot}^2+z^2})
\int\limits_{0}^{\sqrt{R^2-r_{\bot}^2}-z}dl
\rho(\sqrt{r_{\bot}^2+(z+l)^2})
\end{equation}
$$
\times
\exp{\left[-\sigma_{pN}^{\rm in}A\int\limits_{-\sqrt{R^2-r_{\bot}^2}}^{z}\rho(\sqrt{r_{\bot}^2+x^2})dx
-\sigma_{{\pi^{\prime}}N}^{\rm tot}A\int\limits_{z}^{z+l}
\rho(\sqrt{r_{\bot}^2+x^2})dx\right]}
$$
$$
\times
\exp{\left[-\sigma_{K^+N}^{\rm tot}A\int\limits_{z+l}^{\sqrt{R^2-r_{\bot}^2}}
\rho(\sqrt{r_{\bot}^2+x^2})dx
-\sigma_{{\Lambda}N}^{\rm tot}(p^{\prime}_{\Lambda})A
\int\limits_{z+l}^{\sqrt{R^2-r_{\bot}^2}}\rho(\sqrt{r_{\bot}^2+x^2})dx\right]}
$$
and
\begin{equation}
\left<\frac{d\sigma_{{\pi^{\prime}}N \to {\Lambda}K^+X}({\bf p}_{\pi},
{\bf p}^{\prime}_{\Lambda},{\bf p}^{\prime}_{K^+})}{d{\bf p}^{\prime}_{\Lambda}d{\bf p}^{\prime}_{K^+}}\right>_A
=
\int\int
P_A({\bf p}_t,E)d{\bf p}_tdE
\end{equation}
$$
\times
\left\{\frac{d\sigma_{{\pi^{\prime}}N \to {\Lambda}K^+X}[\sqrt{s_1},<m^*_{K^+}>,<m^*_{\Lambda}>,
{\bf p}^{\prime}_{\Lambda},{\bf p}^{\prime}_{K^+}]}{d{\bf p}^{\prime}_{\Lambda}d{\bf p}^{\prime}_{K^+}}\right\}.
$$
Here, $d\sigma_{pA \to {\pi^{\prime}}X}^{\rm (prim)}({\bf p}_0)/d{\bf p}_{\pi}$ are the inclusive differential cross sections for pion production on nuclei at small laboratory angles and for high momenta from the primary proton-induced
reaction channel (48);
$d\sigma_{{\pi^{\prime}}N \to {\Lambda}K^+X}[\sqrt{s_1},<m^*_{K^+}>,<m^*_{\Lambda}>,
{\bf p}^{\prime}_{\Lambda},{\bf p}^{\prime}_{K^+}]/d{\bf p}^{\prime}_{\Lambda}d{\bf p}^{\prime}_{K^+}$
are the ``in-medium" differential cross sections for ${\Lambda}$ and $K^+$ creation with effective masses
$<m^*_{\Lambda}>$ and $<m^*_{K^+}>$ and with the in-medium momenta ${\bf p}^{\prime}_{\Lambda}$ and ${\bf p}^{\prime}_{K^+}$, respectively, in reactions (49) and (50) at the ${\pi^{\prime}}N$ center-of-mass energy $\sqrt{s_1}$. This and the other quantities in Eq.~(51) and (52)
are defined in Refs.~[44, 57] as well as by Eqs.~(29) and (42)--(45). For the cross sections
$d\sigma_{pA \to {\pi^{\prime}}X}^{\rm (prim)}({\bf p}_0)/d{\bf p}_{\pi}$, we use the respective parametrizations
of the experimental pion yields at small angles and for high momenta [44] in our calculations
of the ${\Lambda}K^+$ cross sections from the two-step processes (48)--(50). For $^{63}$Cu target nucleus they
were taken from Ref.~[44]. For ${\Lambda}K^+$ production calculations in the cases of $^{12}$C and $^{108}$Ag,
$^{197}$Au target nuclei presented below, we have supposed that the ratio  of the differential cross section
for pion creation on $^{12}$C and on these nuclei from the primary process (48) to the effective number of
target nucleons participating in it (quantity $I_V^{\prime}[A]$ in Eq.~(51)) is the same as that for $^{9}$Be and
$^{63}$Cu adjusted for kinematics, relating, respectively, to $^{12}$C and $^{108}$Ag, $^{197}$Au.
For the $^{9}$Be target nucleus the parametrizations of the experimental pion differential cross sections
at small angles and for high momenta were borrowed from Ref.~[44].
Within the representation of Eqs.~(49) and (50), the cross sections
$d\sigma_{{\pi^{\prime}}N \to {\Lambda}K^+X}[\sqrt{s_1},<m^*_{K^+}>,<m^*_{\Lambda}>,
{\bf p}^{\prime}_{\Lambda},{\bf p}^{\prime}_{K^+}]/d{\bf p}^{\prime}_{\Lambda}d{\bf p}^{\prime}_{K^+}$
can be written in the following forms:
\begin{equation}
\frac{d\sigma_{{\pi^{+}}p \to {\Lambda}K^+X}[\sqrt{s_1},<m^*_{K^+}>,<m^*_{\Lambda}>,
{\bf p}^{\prime}_{\Lambda},{\bf p}^{\prime}_{K^+}]}{d{\bf p}^{\prime}_{\Lambda}d{\bf p}^{\prime}_{K^+}}=
\frac{d\sigma_{{\pi^{+}}p \to {\Lambda}{\pi^+}K^+}[\sqrt{s_1},<m^*_{K^+}>,<m^*_{\Lambda}>,
{\bf p}^{\prime}_{\Lambda},{\bf p}^{\prime}_{K^+}]}{d{\bf p}^{\prime}_{\Lambda}d{\bf p}^{\prime}_{K^+}},
\end{equation}
\begin{equation}
\frac{d\sigma_{{\pi^{0}}p \to {\Lambda}K^+X}[\sqrt{s_1},<m^*_{K^+}>,<m^*_{\Lambda}>,
{\bf p}^{\prime}_{\Lambda},{\bf p}^{\prime}_{K^+}]}{d{\bf p}^{\prime}_{\Lambda}d{\bf p}^{\prime}_{K^+}}=
\frac{d\sigma_{{\pi^{0}}p \to {\Lambda}K^+}[\sqrt{s_1},<m^*_{K^+}>,<m^*_{\Lambda}>,
{\bf p}^{\prime}_{\Lambda},{\bf p}^{\prime}_{K^+}]}{d{\bf p}^{\prime}_{\Lambda}d{\bf p}^{\prime}_{K^+}}
\end{equation}
$$
+
\frac{d\sigma_{{\pi^{0}}p \to {\Lambda}{\pi^0}K^+}[\sqrt{s_1},<m^*_{K^+}>,<m^*_{\Lambda}>,
{\bf p}^{\prime}_{\Lambda},{\bf p}^{\prime}_{K^+}]}{d{\bf p}^{\prime}_{\Lambda}d{\bf p}^{\prime}_{K^+}},
$$
\begin{equation}
\frac{d\sigma_{{\pi^{-}}p \to {\Lambda}K^+X}[\sqrt{s_1},<m^*_{K^+}>,<m^*_{\Lambda}>,
{\bf p}^{\prime}_{\Lambda},{\bf p}^{\prime}_{K^+}]}{d{\bf p}^{\prime}_{\Lambda}d{\bf p}^{\prime}_{K^+}}=
\frac{d\sigma_{{\pi^{-}}p \to {\Lambda}{\pi^-}K^+}[\sqrt{s_1},<m^*_{K^+}>,<m^*_{\Lambda}>,
{\bf p}^{\prime}_{\Lambda},{\bf p}^{\prime}_{K^+}]}{d{\bf p}^{\prime}_{\Lambda}d{\bf p}^{\prime}_{K^+}},
\end{equation}
\begin{equation}
\frac{d\sigma_{{\pi^{+}}n \to {\Lambda}K^+X}[\sqrt{s_1},<m^*_{K^+}>,<m^*_{\Lambda}>,
{\bf p}^{\prime}_{\Lambda},{\bf p}^{\prime}_{K^+}]}{d{\bf p}^{\prime}_{\Lambda}d{\bf p}^{\prime}_{K^+}}=
\frac{d\sigma_{{\pi^{+}}n \to {\Lambda}K^+}[\sqrt{s_1},<m^*_{K^+}>,<m^*_{\Lambda}>,
{\bf p}^{\prime}_{\Lambda},{\bf p}^{\prime}_{K^+}]}{d{\bf p}^{\prime}_{\Lambda}d{\bf p}^{\prime}_{K^+}}
\end{equation}
$$
+
\frac{d\sigma_{{\pi^{+}}n \to {\Lambda}{\pi^0}K^+}[\sqrt{s_1},<m^*_{K^+}>,<m^*_{\Lambda}>,
{\bf p}^{\prime}_{\Lambda},{\bf p}^{\prime}_{K^+}]}{d{\bf p}^{\prime}_{\Lambda}d{\bf p}^{\prime}_{K^+}},
$$
\begin{equation}
\frac{d\sigma_{{\pi^{0}}n \to {\Lambda}K^+X}[\sqrt{s_1},<m^*_{K^+}>,<m^*_{\Lambda}>,
{\bf p}^{\prime}_{\Lambda},{\bf p}^{\prime}_{K^+}]}{d{\bf p}^{\prime}_{\Lambda}d{\bf p}^{\prime}_{K^+}}=
\frac{d\sigma_{{\pi^{0}}n \to {\Lambda}{\pi^-}K^+}[\sqrt{s_1},<m^*_{K^+}>,<m^*_{\Lambda}>,
{\bf p}^{\prime}_{\Lambda},{\bf p}^{\prime}_{K^+}]}{d{\bf p}^{\prime}_{\Lambda}d{\bf p}^{\prime}_{K^+}},
\end{equation}
\begin{equation}
\frac{d\sigma_{{\pi^{-}}n \to {\Lambda}K^+X}[\sqrt{s_1},<m^*_{K^+}>,<m^*_{\Lambda}>,
{\bf p}^{\prime}_{\Lambda},{\bf p}^{\prime}_{K^+}]}{d{\bf p}^{\prime}_{\Lambda}d{\bf p}^{\prime}_{K^+}}=0,
\end{equation}
where
$d\sigma_{{\pi^{\prime}}N \to {\Lambda}K^+}[\sqrt{s_1},<m^*_{K^+}>,<m^*_{\Lambda}>,
{\bf p}^{\prime}_{\Lambda},{\bf p}^{\prime}_{K^+}]/d{\bf p}^{\prime}_{\Lambda}d{\bf p}^{\prime}_{K^+}$
and
$d\sigma_{{\pi^{\prime}}N \to {\Lambda}{\pi}K^+}[\sqrt{s_1},<m^*_{K^+}>,<m^*_{\Lambda}>,
{\bf p}^{\prime}_{\Lambda},{\bf p}^{\prime}_{K^+}]/d{\bf p}^{\prime}_{\Lambda}d{\bf p}^{\prime}_{K^+}$
are the ``in-medium" differential cross sections of reaction channels (49) and (50), correspondingly.
Taking into account the two-body kinematics of elementary processes (49) as well as the isospin symmetry,
we get the following expressions for the former ones:
\begin{equation}
\frac{d\sigma_{{\pi^{+}}n \to {\Lambda}K^+}[\sqrt{s_1},<m^*_{K^+}>,<m^*_{\Lambda}>,
{\bf p}^{\prime}_{\Lambda},{\bf p}^{\prime}_{K^+}]}{d{\bf p}^{\prime}_{\Lambda}d{\bf p}^{\prime}_{K^+}}=
\frac{\pi}{I_2(s_1,<m^*_{K^+}>,<m^*_{\Lambda}>)E^{\prime}_{\Lambda}}
\end{equation}
$$
\times
\frac{d\sigma_{{\pi^{+}}n \to {\Lambda}K^+}(\sqrt{s_1},<m^*_{K^+}>,<m^*_{\Lambda}>,\theta^*_{\Lambda})}
{d{\bf \Omega}^*_{\Lambda}}
$$
$$
\times
\frac{1}{(\omega_0+E_t)}\delta\left[\omega_0+E_t-\sqrt{(<m^*_{K^+}>)^2+{\bf p}^{{\prime}2}_{K^+}}\right]
\delta({\bf Q}_0+{\bf p}_t-{\bf p}^{\prime}_{K^+}),
$$
\begin{equation}
\frac{d\sigma_{{\pi^{0}}p \to {\Lambda}K^+}[\sqrt{s_1},<m^*_{K^+}>,<m^*_{\Lambda}>,
{\bf p}^{\prime}_{\Lambda},{\bf p}^{\prime}_{K^+}]}{d{\bf p}^{\prime}_{\Lambda}d{\bf p}^{\prime}_{K^+}}=
\frac{1}{2}\frac{d\sigma_{{\pi^{+}}n \to {\Lambda}K^+}[\sqrt{s_1},<m^*_{K^+}>,<m^*_{\Lambda}>,
{\bf p}^{\prime}_{\Lambda},{\bf p}^{\prime}_{K^+}]}{d{\bf p}^{\prime}_{\Lambda}d{\bf p}^{\prime}_{K^+}}.
\end{equation}
Here,
\begin{equation}
\omega_0=E_{\pi}-E^{\prime}_{\Lambda},\,\,\,
{\bf Q}_0={\bf p}_{\pi}-{\bf p}^{\prime}_{\Lambda},
\end{equation}
where ${\bf p}_{\pi}$ and $E_{\pi}$ are the momentum and total energy respectively of an intermediate pion
(which is assumed to be on-shell), and $I_2$ is the two-body phase space, defined as [58]:
\begin{equation}
I_{2}(s_1,<m^*_{K^+}>,<m^*_{\Lambda}>)=(\frac{{\pi}}{2})
\frac{\lambda[s_1,(<m^*_{K^+}>)^{2},(<m^*_{\Lambda}>)^{2}]}{s_1}.
\end{equation}
In Eq.~(60),
$d\sigma_{{\pi^{+}}n \to {\Lambda}K^+}(\sqrt{s_1},<m^*_{K^+}>,<m^*_{\Lambda}>,\theta^*_{\Lambda})/d{\bf \Omega}^*_{\Lambda}$
is the $\Lambda$ ``in-medium" differential cross section in the $\pi^+n$ center-of-mass system.
As earlier, we assume that this cross section is equivalent to the vacuum cross section
$d\sigma_{{\pi^{+}}n \to {\Lambda}K^+}(\sqrt{s_1},m_{K^+},m_{\Lambda},\theta^*_{\Lambda})/d{\bf \Omega}^*_{\Lambda}$ in which the free kaon and $\Lambda$ hyperon masses $m_{K^+}$ and $m_{\Lambda}$ are
replaced by the in-medium masses $<m^*_{K^+}>$ and $<m^*_{\Lambda}>$. As in Refs.~[60, 65], we choose the free $\Lambda$ angular distribution in the following form:
\begin{equation}
\frac{d\sigma_{{\pi^{+}}n \to {\Lambda}K^+}(\sqrt{s_1},m_{K^+},m_{\Lambda},\theta^*_{\Lambda})}
{d{\bf \Omega}^*_{\Lambda}}=[1-A_1(\sqrt{s_1},\sqrt{{\tilde s}_{{\rm th}}})\cos{\theta^*_{\Lambda}}]
\frac{\sigma_{{\pi^{+}}n \to {\Lambda}K^+}(\sqrt{s_1},\sqrt{{\tilde s}_{{\rm th}}})}{4\pi},
\end{equation}
\begin{equation}
A_1({\sqrt{s_1},\sqrt{{\tilde s}_{{\rm th}}}})=\left\{
\begin{array}{ll}
	5.26\left(\frac{\sqrt{s_1}-\sqrt{{\tilde s}_{{\rm th}}}}{\rm GeV}\right)
	&\mbox{for $\sqrt{{\tilde s}_{{\rm th}}} < \sqrt{s_1} \le 1.8~{\rm GeV}$}, \\
	&\\
                   1
	&\mbox{for $\sqrt{s_1} > 1.8~{\rm GeV}$},
\end{array}
\right.	
\end{equation}
\begin{equation}
\sigma_{{\pi^+}n \to {\Lambda}K^+}({\sqrt{s_1},\sqrt{{\tilde s}_{{\rm th}}}})=\left\{
\begin{array}{ll}
	10.0\left(\frac{\sqrt{s_1}-\sqrt{{\tilde s}_{{\rm th}}}}{\rm GeV}\right)~[{\rm mb}]
	&\mbox{for $\sqrt{{\tilde s}_{{\rm th}}} < \sqrt{s_1} \le 1.7~{\rm GeV}$}, \\
	&\\
                   0.09\left(\frac{{\rm GeV}}{\sqrt{s_1}-1.6~{\rm GeV}}\right)~[{\rm mb}]
	&\mbox{for $\sqrt{s_1} > 1.7~{\rm GeV}$},
\end{array}
\right.	
\end{equation}
where $\sqrt{{\tilde s}_{{\rm th}}}=m_{\Lambda}+m_{K^+}$ is the free threshold energy.
In our calculations of the ${\Lambda}K^+$ pair production on nuclei, the ``in-medium" differential cross sections
$d\sigma_{{\pi^{\prime}}N \to {\Lambda}{\pi}K^+}[\sqrt{s_1},<m^*_{K^+}>,<m^*_{\Lambda}>,
{\bf p}^{\prime}_{\Lambda},{\bf p}^{\prime}_{K^+}]/d{\bf p}^{\prime}_{\Lambda}d{\bf p}^{\prime}_{K^+}$
of reaction channels (50) have been described according to the three-body phase space. Following Eq.~(36),
one has:
\begin{equation}
\frac{d\sigma_{{\pi^{\prime}}N\to {\Lambda}{\pi}K^+}[\sqrt{s_1},<m^*_{K^+}>,<m^*_{\Lambda}>,
{\bf p}^{\prime}_{\Lambda},{\bf p}^{\prime}_{K^+}]}{d{\bf p}^{\prime}_{\Lambda}d{\bf p}^{\prime}_{K^+}}=
\frac{1}{8E^{\prime}_{\Lambda}E^{\prime}_{K^+}}
\end{equation}
$$
\times
\frac{\sigma_{{\pi^{\prime}}N\to {\Lambda}{\pi}K^+}(\sqrt{s_1},
\sqrt{{\tilde s}^*_{1{\rm th}}})}{I_3(s_1,<m^*_{K^+}>,m_{\pi},<m^*_{\Lambda}>)}
\frac{1}{(\omega_1+E_t)}
\delta\left(\omega_1+E_t-
\sqrt{m_{\pi}^2+({\bf Q}_1+{\bf p}_t)^2}\right),
$$
where
\begin{equation}
\omega_1=E_{\pi}-E^{\prime}_{\Lambda}-E^{\prime}_{K^+},\,\,\,
{\bf Q}_1={\bf p}_{\pi}-{\bf p}^{\prime}_{\Lambda}-{\bf p}^{\prime}_{K^+}
\end{equation}
and $\sqrt{{\tilde s}^*_{1{\rm th}}}=<m^*_{\Lambda}>+m_{\pi}+<m^*_{K^+}>$ is the in-medium threshold energy. In Eq.~(67),\\
$\sigma_{{\pi^{\prime}}N \to {\Lambda}{\pi}K^+}(\sqrt{s_1},\sqrt{{\tilde s}^*_{1{\rm th}}})$
are the ``in-medium" total cross sections for ${\Lambda}K^+$ pair production in the reactions (50).
As before, we assume that these cross sections are equivalent to the vacuum cross sections\\
$\sigma_{{\pi^{\prime}}N \to {\Lambda}{\pi}K^+}(\sqrt{s_1},\sqrt{{\tilde s}_{1{\rm th}}})$
in which the free threshold energy $\sqrt{{\tilde s}_{1{\rm th}}}=m_{\Lambda}+m_{\pi}+m_{K^+}$
is replaced by the in-medium threshold $\sqrt{{\tilde s}^*_{1{\rm th}}}$.
In line with Ref.~[35], for the free total cross sections
$\sigma_{{\pi^{\prime}}N \to {\Lambda}{\pi}K^+}(\sqrt{s_1},\sqrt{{\tilde s}_{1{\rm th}}})$
we have adopted the following expression:
\begin{equation}
\sigma_{{\pi^+}p \to {\Lambda}{\pi^+}K^+}(\sqrt{s_1},\sqrt{{\tilde s}_{1{\rm th}}})
\approx \sigma_{{\pi^0}p \to {\Lambda}{\pi^0}K^+}(\sqrt{s_1},\sqrt{{\tilde s}_{1{\rm th}}})
\approx \sigma_{{\pi^-}p \to {\Lambda}{\pi^-}{K^+}}(\sqrt{s_1},\sqrt{{\tilde s}_{1{\rm th}}})
\end{equation}
$$
\approx \sigma_{{\pi^+}n \to {\Lambda}{\pi^0}K^+}(\sqrt{s_1},\sqrt{{\tilde s}_{1{\rm th}}})
\approx \sigma_{{\pi^0}n \to {\Lambda}{\pi^-}K^+}(\sqrt{s_1},\sqrt{{\tilde s}_{1{\rm th}}})
\approx 24.0\left(1-\frac{{\tilde s}_{1{\rm th}}}{s_1}\right)^{3.16}
\left(\frac{{\tilde s}_{1{\rm th}}}{s_1}\right)^{4.24}~[\rm mb].
$$
It is worth mentioning that, as follows from Eqs.~(66), (69) and
Ref.~[66], the ${\Lambda}K^+$ production cross sections in the secondary pion--nucleon
processes (49) and (50) are substantially larger than those in the four-body reaction channels
${\pi^{\prime}}N \to {\Lambda}{\pi}{\pi}K^+$ at pion momenta $\le$ 2 GeV/c, giving, as our estimate shows, the main contribution to the ${\Lambda}K^+$ creation on nuclei for kinematics
of interest. Therefore, we discard the latter in the present study.

   In addition to the two-step processes with intermediate pions (48)--(50) we consider the
following production/decay sequence, which may contribute to the ${\Lambda}K^+$ yield from nuclei
in the conditions of the ANKE experiment at the incident proton beam energy of interest:
\begin{eqnarray}
p+p \to \Sigma^0+p+K^+,
\end{eqnarray}
\begin{eqnarray}
p+n \to \Sigma^0+n+K^+,
\end{eqnarray}
\begin{eqnarray}
\Sigma^0 \to \Lambda+\gamma.
\end{eqnarray}
Presently, there are four sets of data  available for the total cross section $\sigma_{pp \to {\Sigma^0}pK^+}$
of reaction (70). Three of these have recently been  taken by the COSY-11 [24, 25],
COSY-TOF [28] and ANKE [29] Collaborations at proton energies $\le$ 2.4 GeV,
whereas the other was obtained a long time ago at beam energies
$\ge$ 2.85 GeV [21]. The comparison of these data with the results of calculations by parametrization
\begin{equation}
\sigma_{pp \to {\Sigma^0}pK^+}(\sqrt{s},\sqrt{s_{0}})=\left\{
\begin{array}{ll}
	\frac{A_{\Sigma^0}(s-s_{0})^2}{4m_p^2+B_{\Sigma^0}(s-s_{0})^2}
	&\mbox{for $0.225~{\rm GeV} < \sqrt{s}-\sqrt{s_{0}} < 2.0~{\rm GeV}$}, \\
	&\\
       C_{\Sigma^0}\left(\sqrt{s}-\sqrt{s_{0}}\right)^2
	&\mbox{for $0 < \sqrt{s}-\sqrt{s_{0}} \le 0.225~{\rm GeV}$},
\end{array}
\right.	
\end{equation}
(solid line) is shown in Fig.~4. In Eq.~(73),
$\sqrt{s_{0}}=m_{\Sigma^0}+m_p+m_{K^+}$ is the threshold energy and the constants
$A_{\Sigma^0}$, $B_{\Sigma^0}$ and $C_{\Sigma^0}$ are given as 26.0 $\mu$b/GeV$^2$, 1/GeV$^2$ and
154.5 $\mu$b/GeV$^2$, respectively. The ``low" excess energy part of Eq.~(73) was taken from Ref.~[28]. The parametrization (73) fits well the full available set of data
for the $pp \to {\Sigma^0}pK^+$ process
\footnote{$^)$This parametrization is also consistent with the cross
section of (16.5$\pm$20\%)~${\rm {\mu}b}$ for channel $pp \to {\Sigma^0}pK^+$ at beam energy of 3.5 GeV
($\sqrt{s}-\sqrt{s_0}=0.555$ GeV) evaluated in Ref.~[30], by dividing the measured cross section for the
$pp \to {\Lambda}pK^+$ reaction at this energy by a factor of 2.2, and assuming an uncertainty of 20\%.}$^)$
.
\begin{figure}[!h]
\begin{center}
\includegraphics[width=12.0cm]{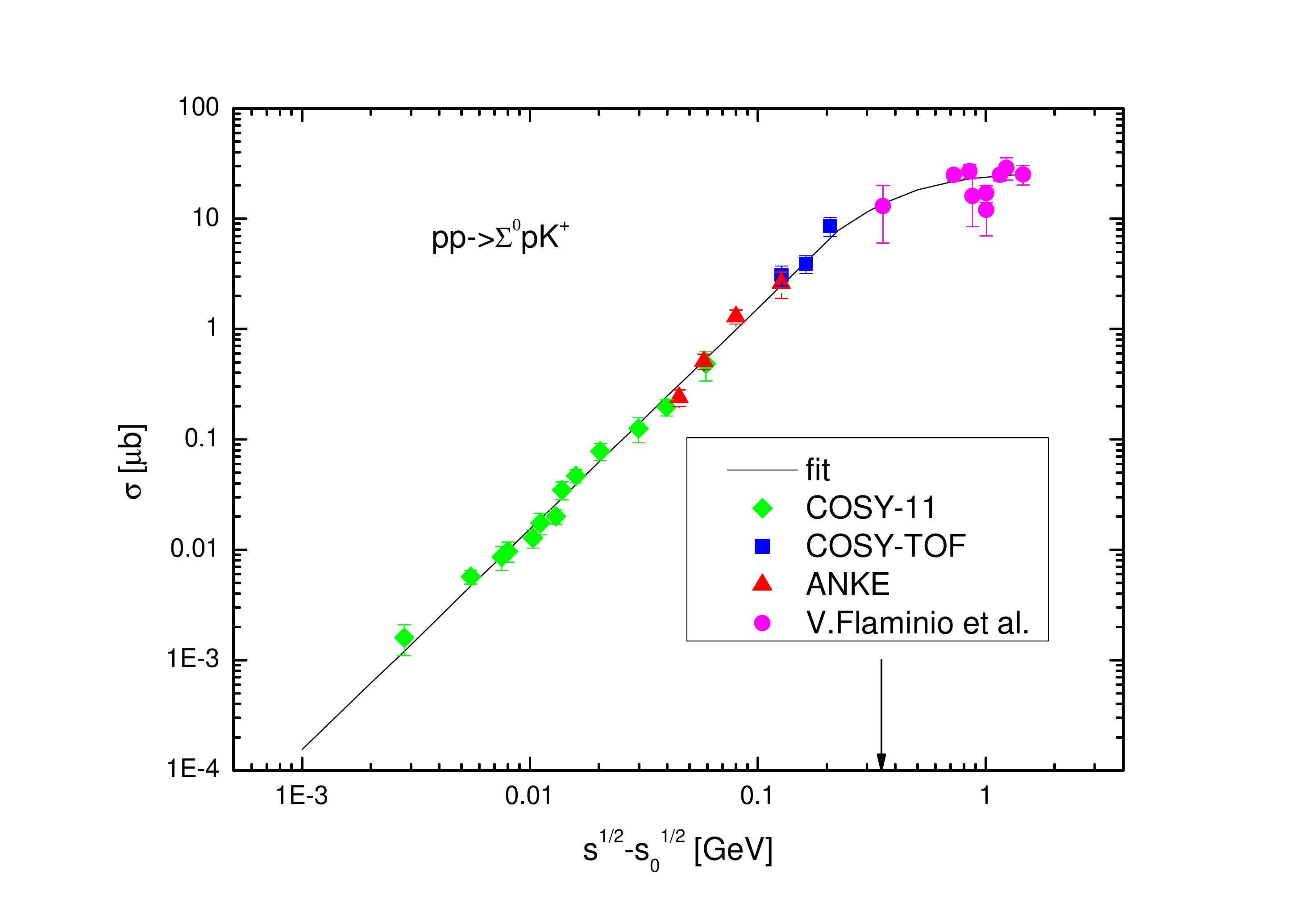}
\vspace*{-2mm} \caption{Total cross section for the $pp \to {\Sigma^0}pK^+$ reaction
as a function of excess energy. The arrow indicates the excess energy, which corresponds
to the proton kinetic energy of 2.83 GeV. For further notation see the text.}
\label{void}
\end{center}
\end{figure}
Direct data on the total cross section $\sigma_{pn \to {\Sigma^0}nK^+}$ of reaction (71) do not currently
exist. Reference~[21] only has three data points for the total cross section
$\sigma_{pn \to {\Sigma^0}pK^0}$ of the channel $pn \to {\Sigma^0}pK^0$ at 5.135, 6.124
and 16.088 GeV initial energies, which due to the isospin symmetry is equal to the former one.
Comparing these with the data that are available [21] for the $pp \to {\Sigma^0}pK^+$ process at similar
energies (at energies of 5.135 and 6.045 GeV), one can readily find that at these energies the ratio
of the $pn$ and $pp$ total cross sections $\sigma_{pn \to {\Sigma^0}pK^0}/\sigma_{pp \to {\Sigma^0}pK^+}$
varies approximately from 0.5 to 1.5. Therefore, it is natural to take for this ratio here an average
value of one, which means that:
\begin{equation}
\sigma_{pn \to {\Sigma^0}pK^0}(\sqrt{s})=\sigma_{pn \to {\Sigma^0}nK^+}(\sqrt{s})
\approx \sigma_{pp \to {\Sigma^0}pK^+}(\sqrt{s},\sqrt{s_{0}})
\end{equation}
at the  high incident proton kinetic energies considered. Due to the lack of   data
at lower beam energies, we will also adopt Eq.~(74)  at all collision energies $\sqrt{s}$,
accessible in the calculation of ${\Lambda}K^+$ production in $pA$ interactions from the production/decay
sequence (70)--(72) at the initial energy of 2.83 GeV. At this energy the ratio  $R_{\Sigma^0/\Lambda}$
of the total cross sections $\sigma_{pp \to {\Sigma^0}pK^+}$ and $\sigma_{pp \to {\Lambda}pK^+}$, shown in
Figs.~ 4 and 1, is about 1/4. Hence, the production/decay chain (70)--(72) may indeed contribute to the
$(p,{\Lambda}K^+)$ reaction on nuclei for our initial energy of  interest. The total
cross section of the subprocess $pp \to {\Sigma^0}p{\pi^0}K^+$ with additional pion in the final state, assuming that it goes completely through the reaction $pp \to {\Lambda}(1405)pK^+$, is (1.5$\pm$0.7) ${\rm \mu}$b
at beam energy of 2.83 GeV [67]. This cross section is substantially lower than that for the
$pp \to {\Sigma^0}pK^+$ process, shown in Fig. 4, at this energy. Therefore, we will neglect the
contribution from the production/decay chain $pN \to {\Sigma^0}N{\pi^0}K^+$,
${\Sigma^0} \to {\Lambda}{\gamma}$ in our calculations of the ${\Lambda}K^+$ yield in $pA$ collisions
at 2.83 GeV incident energy.

The  $\Sigma^0$ hyperons and $K^+$ mesons produced in elementary processes (70), (71) are mainly emitted in the
forward direction.  Most of the $\Sigma^0$'s decay into $\Lambda$ and $\gamma$ essentially outside the
target nuclei of interest. Taking into account these facts
and neglecting the change of the $\Sigma^0$ mass in the nuclear medium but
accounting for the in-medium modifications of the masses of other final hadrons (kaons and nucleons)
participating in these processes on the same footing as that employed in calculating the ${\Lambda}K^+$
production cross section (26) from the primary proton-induced reaction channels (1)--(6), as well as using the
results given in Refs.~[68, 69], we get the following expression for the ${\Lambda}K^+$ creation cross section
for $pA$ interactions from this chain:
\begin{equation}
\frac{d\sigma_{pA\to {\Lambda}K^+X}^{({\rm sec}),(\Sigma^0)}
({\bf p}_0,{\bf p}_{\Lambda},{\bf p}_{K^+})}
{d{\bf p}_{\Lambda}d{\bf p}_{K^+}}=I_{K^+{\Sigma^0}}[A]\int d{\bf p}_{\Sigma^0}
\end{equation}
$$
\times
\left[\frac{Z}{A}
\left<\frac{d\sigma_{pp\to {\Sigma^0}pK^+}({\bf p}^{\prime}_{0},{\bf p}_{\Sigma^0},
{\bf p}^{\prime}_{K^+})}{d{\bf p}_{\Sigma^0}d{\bf p}^{\prime}_{K^+}}\right>_A+
\frac{N}{A}
\left<\frac{d\sigma_{pn\to {\Sigma^0}nK^+}({\bf p}^{\prime}_{0},
{\bf p}_{\Sigma^0},{\bf p}^{\prime}_{K^+})}{d{\bf p}_{\Sigma^0}d{\bf p}^{\prime}_{K^+}}\right>_A
\right]\frac{d{\bf p}^{\prime}_{K^+}}{d{\bf p}_{K^+}}
$$
$$
\times
\frac{BR(\Sigma^0 \to {\Lambda}\gamma)\theta(E_{\Sigma^0}-E_{\Lambda})}
{4I_2(m_{\Sigma^0}^2,m_{\Lambda}^2,0)E_{\Lambda}(E_{\Sigma^0}-E_{\Lambda})}
\delta\left(E_{\Sigma^0}-E_{\Lambda}-\sqrt{({\bf p}_{\Sigma^0}-{\bf p}_{\Lambda})^2}\right),
$$
where ${\bf p}_{\Sigma^0}$ and $E_{\Sigma^0}$ are the vacuum momentum and total energy of a $\Sigma^0$
hyperon ($E_{\Sigma^0}=\sqrt{m_{\Sigma^0}^2+{\bf p}_{\Sigma^0}^2}$), $\theta(x)$ is the standard step
function and $BR(\Sigma^0 \to {\Lambda}\gamma)=1$.
The averaged differential cross sections\\
$\left<d\sigma_{pN\to {\Sigma^0}NK^+}({\bf p}^{\prime}_{0},{\bf p}_{\Sigma^0},
{\bf p}^{\prime}_{K^+})/d{\bf p}_{\Sigma^0}d{\bf p}^{\prime}_{K^+}\right>_A$,
in Eq.~(75),
are defined by Eqs.~(28), (36) and (37), in which one has to make the following substitutions:
${\bf p}^{\prime}_{\Lambda} \to {\bf p}_{\Sigma^0}$, $E^{\prime}_{\Lambda} \to E_{\Sigma^0}$,
$<m_{\Lambda}^*> \to m_{\Sigma^0}$ and
$\sigma_{pN\to {\Lambda}K^+X}(\sqrt{s},\sqrt{s^*_{\rm th}},
\sqrt{s^*_{1{\rm th}}},\sqrt{s^*_{2{\rm th}}}) \to
\sigma_{pN\to {\Sigma^0}NK^+}(\sqrt{s},\sqrt{s^*_{0}})$,
where $\sqrt{s^*_{0}}=m_{\Sigma^0}+<m_p^*>+<m^*_{K^+}>$. The quantity $I_{K^+{\Sigma^0}}[A]$ in Eq.~(75)
is defined above by Eq.~(27), in which one has to replace $\sigma_{{\Lambda}N}^{\rm tot}$
by inelastic cross section $\sigma_{{\Sigma^0}N}^{\rm in}$ of the ${\Sigma^0}N$ interaction
\footnote{$^)$Using this cross section in (Eq.~27), we assume that the quasi-elastic $\Sigma^0$
rescatterings on intranuclear nucleons do not lead to the loss of $\Sigma^0$ hyperons, which may
undergo subsequently  $\Sigma^0 \to {\Lambda}\gamma$ decays. }$^)$.
Due to isospin symmetry, this cross section is the same as the inelastic cross sections
$\sigma_{{\Sigma^0}p}^{\rm in}$ and $\sigma_{{\Sigma^0}n}^{\rm in}$ of the
${\Sigma^0}p$ and ${\Sigma^0}n$ interactions. At $\Sigma^0$ momenta of interest,
the cross section $\sigma_{{\Sigma^0}p}^{\rm in}$ is exhausted by
the total cross sections $\sigma_{{\Sigma^0}p \to {\Lambda}p}$ and $\sigma_{{\Sigma^0}p \to {\Sigma}^+n}$
of the inelastic ${\Sigma^0}p \to {\Lambda}p$ and ${\Sigma^0}p \to {\Sigma}^+n$ processes:
\begin{equation}
\sigma_{{\Sigma^0}N}^{\rm in}=\sigma_{{\Sigma^0}p}^{\rm in}=\sigma_{{\Sigma^0}p \to {\Lambda}p}
+\sigma_{{\Sigma^0}p \to {\Sigma}^+n}.
\end{equation}
The first cross section in Eq.~(76) is obtained by detailed balance [62]:
\begin{equation}
\sigma_{{\Sigma^0}p \to {\Lambda}p}=\left(\frac{p_{\Lambda}^{\rm cm}}{p_{\Sigma}^{\rm cm}}\right)^2
\sigma_{{\Lambda}p \to {\Sigma}^0p}(p^{\prime}_{\Lambda}),
\end{equation}
where the quantities $\sigma_{{\Lambda}p \to {\Sigma}^0p}$,
$p_{\Lambda}^{\rm cm}$ and $p_{\Sigma}^{\rm cm}$ are defined above by Eqs.~(44), (45), in which
one has to put:
\begin{equation}
s_{\Sigma}=(E_{\Sigma^0}+m_N)^2-p_{\Sigma^0}^2,\,\,
p^{\prime}_{\Lambda}=\sqrt{E^{\prime2}_{\Lambda}-(<m^*_{\Lambda}>)^2},\,\,
E^{\prime}_{\Lambda}=[s_{\Sigma}-m_N^2-(<m^*_{\Lambda}>)^2]/(2m_N).
\end{equation}
For the second cross section in Eq.~(76) we adopt the following parametrization, suggested in Ref.~[62]:
\begin{equation}
\sigma_{{\Sigma^0}p \to {\Sigma^+}n}(p_{\Sigma^0})=22.4/p_{\Sigma^0}-1.08~[{\rm mb}],
\end{equation}
where the $\Sigma^0$ momentum $p_{\Sigma^0}$ is measured in GeV/c.

The differential cross section for $\Lambda$ hyperon production in $pA$ collisions in coincidence
with the $K^+$ meson from the two-step processes (48)--(50) and (70)--(72),
corresponding to the kinematical conditions of the ANKE experiment,
can be defined analogously to Eq.~(46) as:
\begin{equation}
\left<\frac{d\sigma_{pA\to {\Lambda}X}^{({\rm sec})}
({\bf p}_0,p_{\Lambda})}
{dp_{\Lambda}d{\bf \Omega}_{\Lambda}}\right>_
{\Delta{\bf \Omega}_{\Lambda}\Delta{\bf p}_{K^+}}=
\frac{1}{(2\pi)(1-\cos{6^{\circ}})}
\end{equation}
$$
\times
\int\limits_{0.2~{\rm GeV/c}}^{0.6~{\rm GeV/c}}dp_{K^+}
\int\limits_{\cos{12^{\circ}}}^{1}d\cos{\theta_{K^+}}
\int\limits_{\cos{6^{\circ}}}^{1}d\cos{\theta_{\Lambda}}
\int\limits_{0}^{2\pi}d\phi_{K^+}
\int\limits_{0}^{2\pi}d\phi_{\Lambda}
$$
$$
\times
\left[\frac{d\sigma_{pA\to {\Lambda}K^+X}^{({\rm sec}),(\pi)}
({\bf p}_0,{\bf p}_{\Lambda},{\bf p}_{K^+})}
{d{\bf p}_{\Lambda}d{\bf p}_{K^+}}+
\frac{d\sigma_{pA\to {\Lambda}K^+X}^{({\rm sec}),(\Sigma^0)}
({\bf p}_0,{\bf p}_{\Lambda},{\bf p}_{K^+})}
{d{\bf p}_{\Lambda}d{\bf p}_{K^+}}\right]
p^2_{\Lambda}p^2_{K^+}.
$$

We now discuss the results of calculations within the approach outlined above.

\begin{figure}[!h]
\begin{center}
\includegraphics[width=12.0cm]{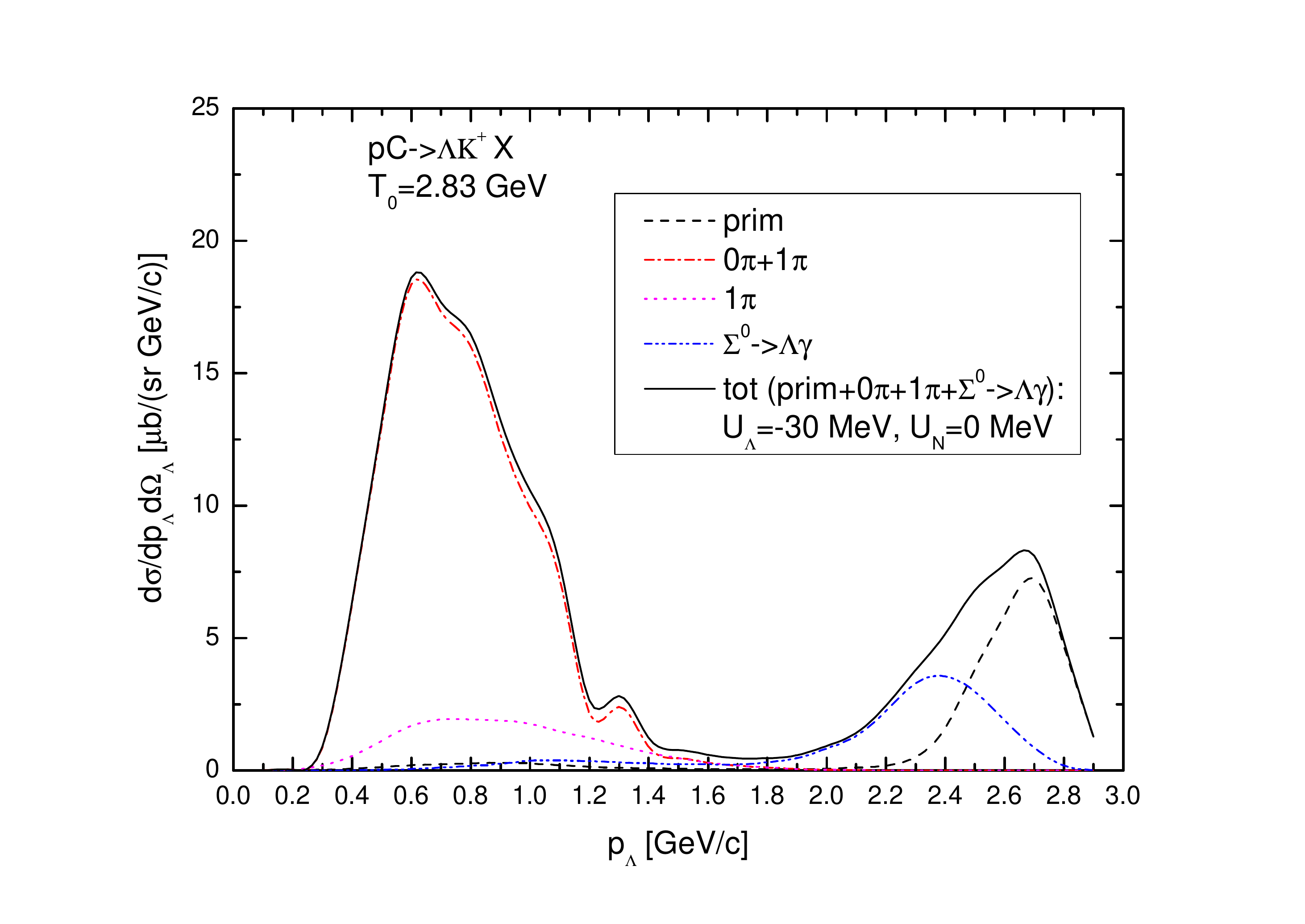}
\vspace*{-2mm} \caption{Differential cross section for the production of $\Lambda$ hyperons
in coincidence with $K^+$ mesons
from primary (1)--(6) (dashed line), secondary (49), (50) (dotted-dashed line),
secondary (50) (dotted line), secondary (72) (dot-dot-dashed line) and primary (1)--(6) plus
secondary (49), (50), (72) (solid line) channels in the ANKE acceptance window as a function of lambda momentum in the interaction of protons of energy of 2.83 GeV with C target nucleus for $\Lambda$
effective scalar potential depth $U_{\Lambda}=-30$ MeV.}
\label{void}
\end{center}
\end{figure}
\begin{figure}[!h]
\begin{center}
\includegraphics[width=12.0cm]{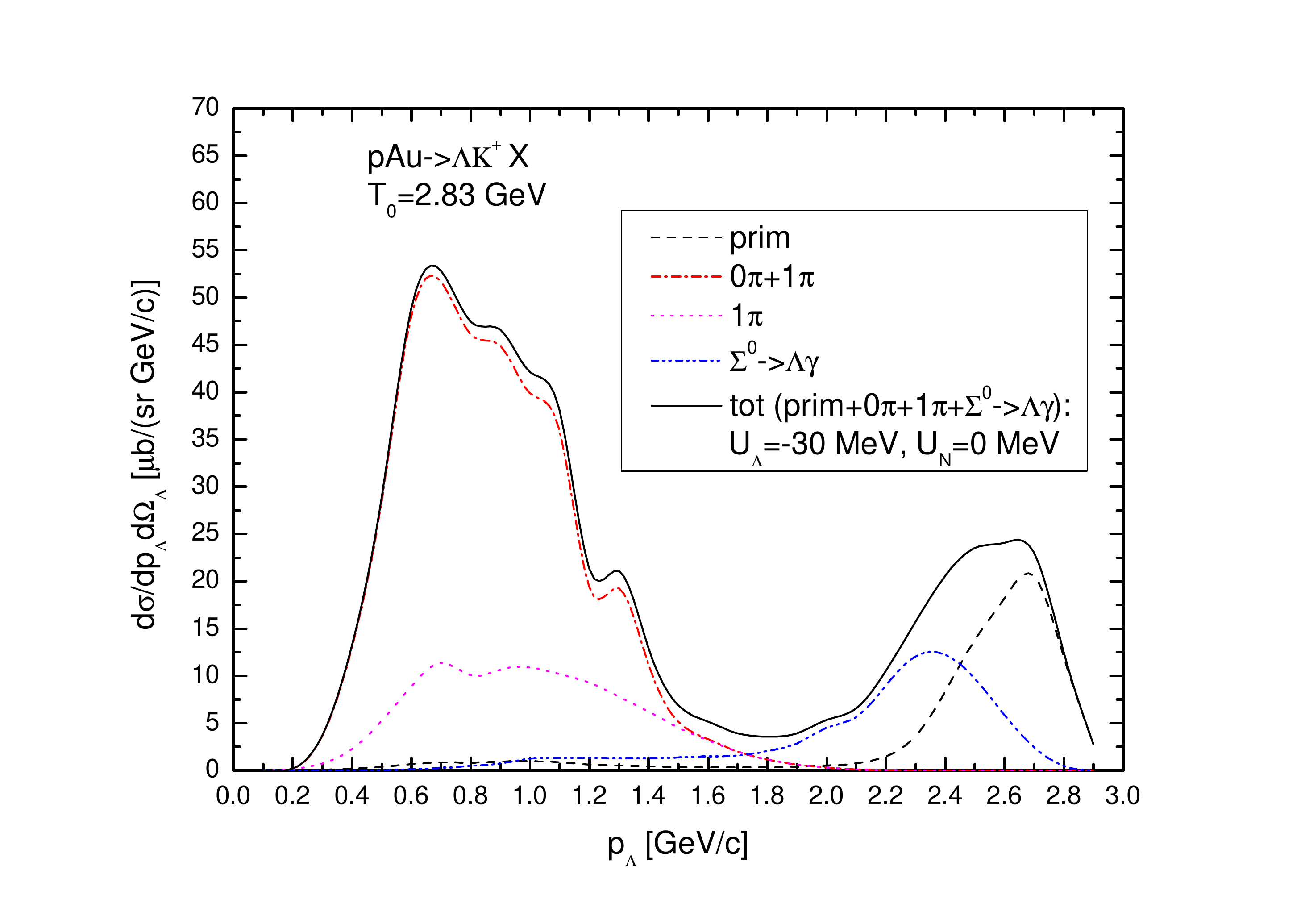}
\vspace*{-2mm} \caption{Differential cross section for the production of $\Lambda$ hyperons
in coincidence with $K^+$ mesons
from primary (1)--(6) (dashed line), secondary (49), (50) (dotted-dashed line),
secondary (50) (dotted line), secondary (72) (dot-dot-dashed line) and primary (1)--(6) plus
secondary (49), (50), (72) (solid line) channels in the ANKE acceptance window as a function of lambda momentum in the interaction of protons of energy of 2.83 GeV with Au target nucleus for $\Lambda$
effective scalar potential depth $U_{\Lambda}=-30$ MeV.}
\label{void}
\end{center}
\end{figure}
\begin{figure}[!h]
\begin{center}
\includegraphics[width=12.0cm]{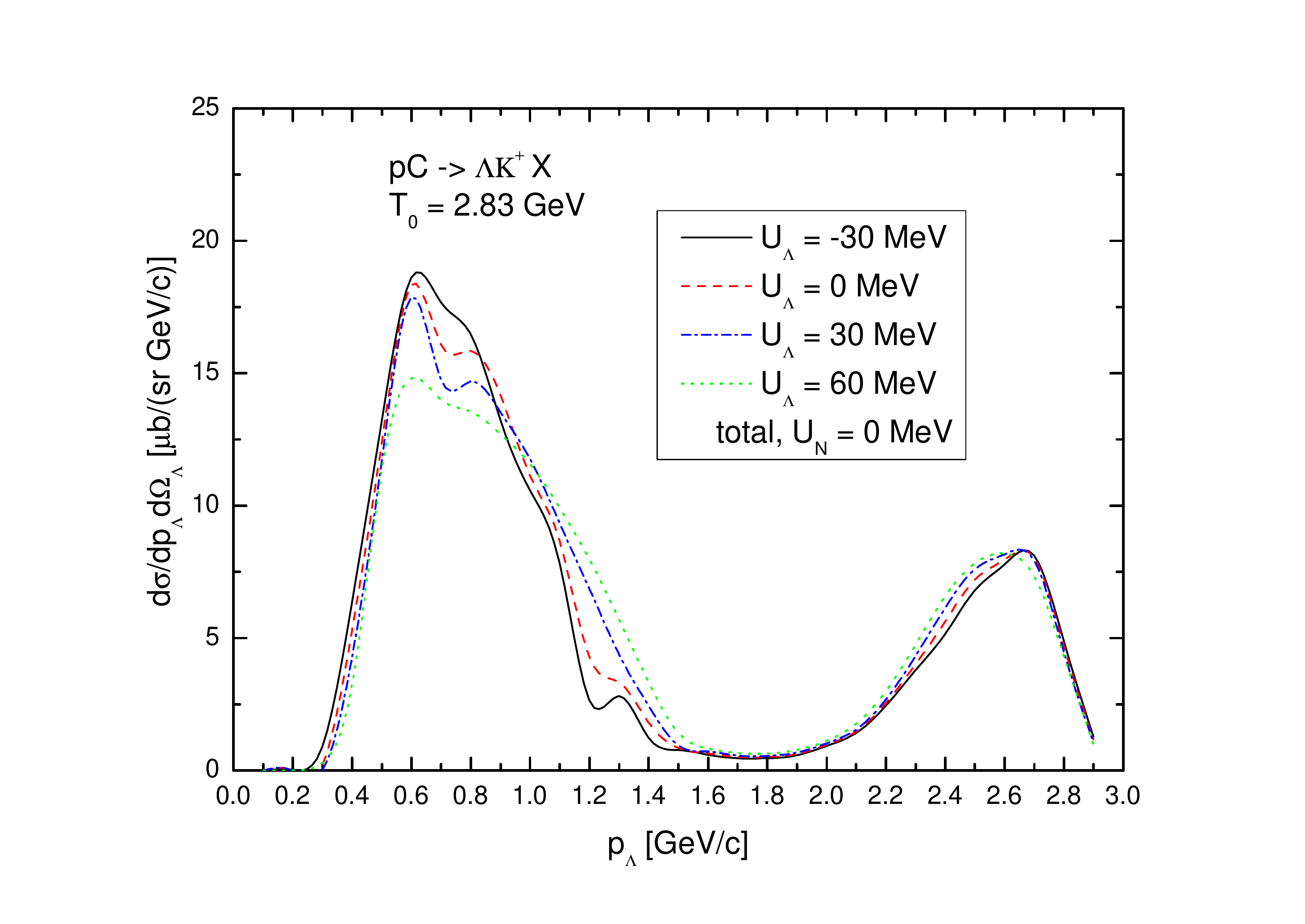}
\vspace*{-2mm} \caption{Differential cross section for the production of $\Lambda$ hyperons
in coincidence with the $K^+$ mesons from primary plus secondary channels
in the ANKE acceptance window as a function of lambda momentum
in the interaction of protons of energy of 2.83 GeV with C target nucleus for effective scalar
$\Lambda$ potentials at saturation density
$U_{\Lambda}=-30$ MeV (solid line), $U_{\Lambda}=0$ MeV (dashed line),
$U_{\Lambda}=30$ MeV (dotted-dashed line) and $U_{\Lambda}=60$ MeV (dotted line).}
\label{void}
\end{center}
\end{figure}
\begin{figure}[!h]
\begin{center}
\includegraphics[width=12.0cm]{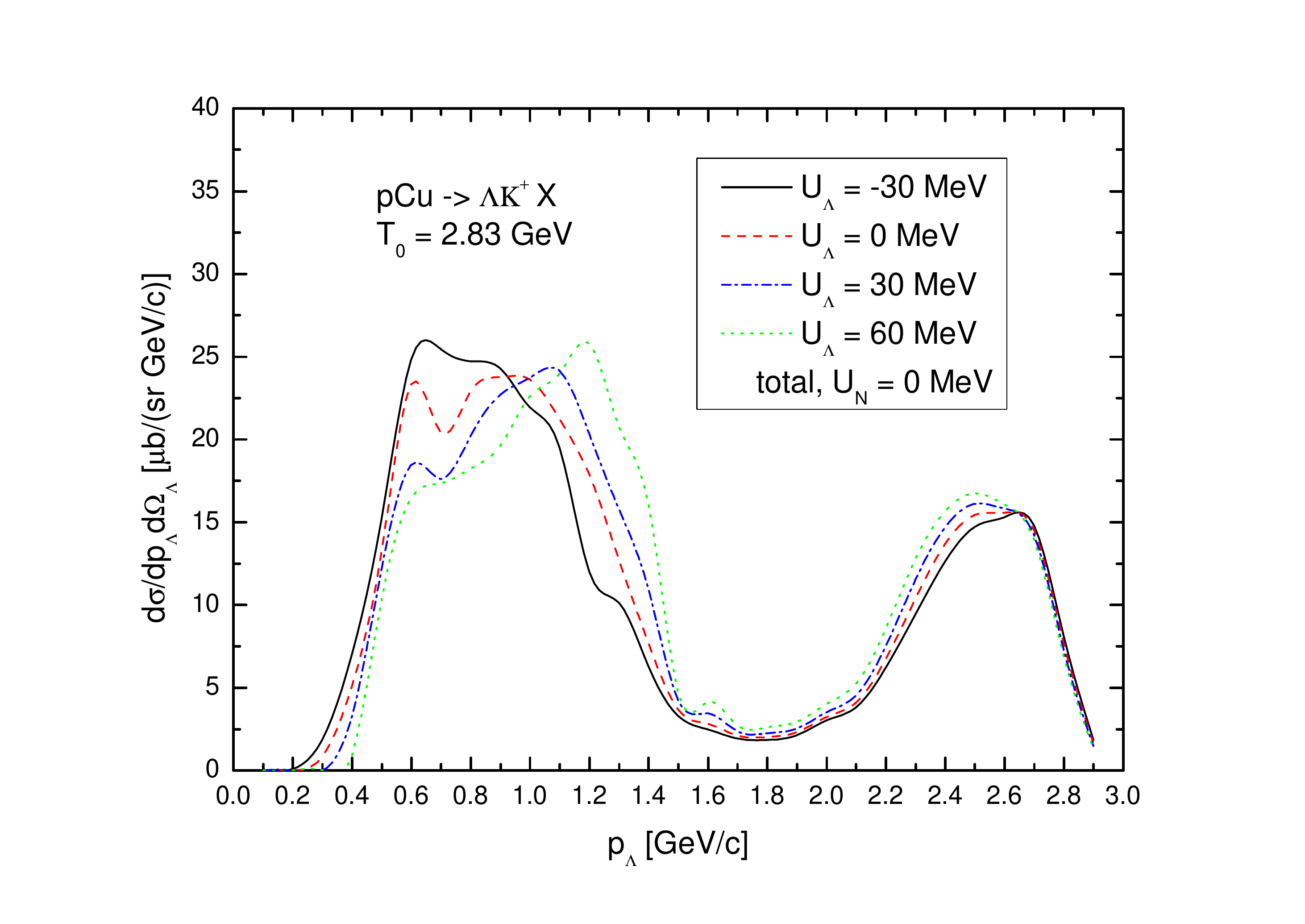}
\vspace*{-2mm} \caption{Differential cross section for the production of $\Lambda$ hyperons
in coincidence with the $K^+$ mesons from primary plus secondary channels
in the ANKE acceptance window as a function of lambda momentum
in the interaction of protons of energy of 2.83 GeV with Cu target nucleus for effective scalar
$\Lambda$ potentials at saturation density
$U_{\Lambda}=-30$ MeV (solid line), $U_{\Lambda}=0$ MeV (dashed line),
$U_{\Lambda}=30$ MeV (dotted-dashed line) and $U_{\Lambda}=60$ MeV (dotted line).}
\label{void}
\end{center}
\end{figure}
\begin{figure}[!h]
\begin{center}
\includegraphics[width=12.0cm]{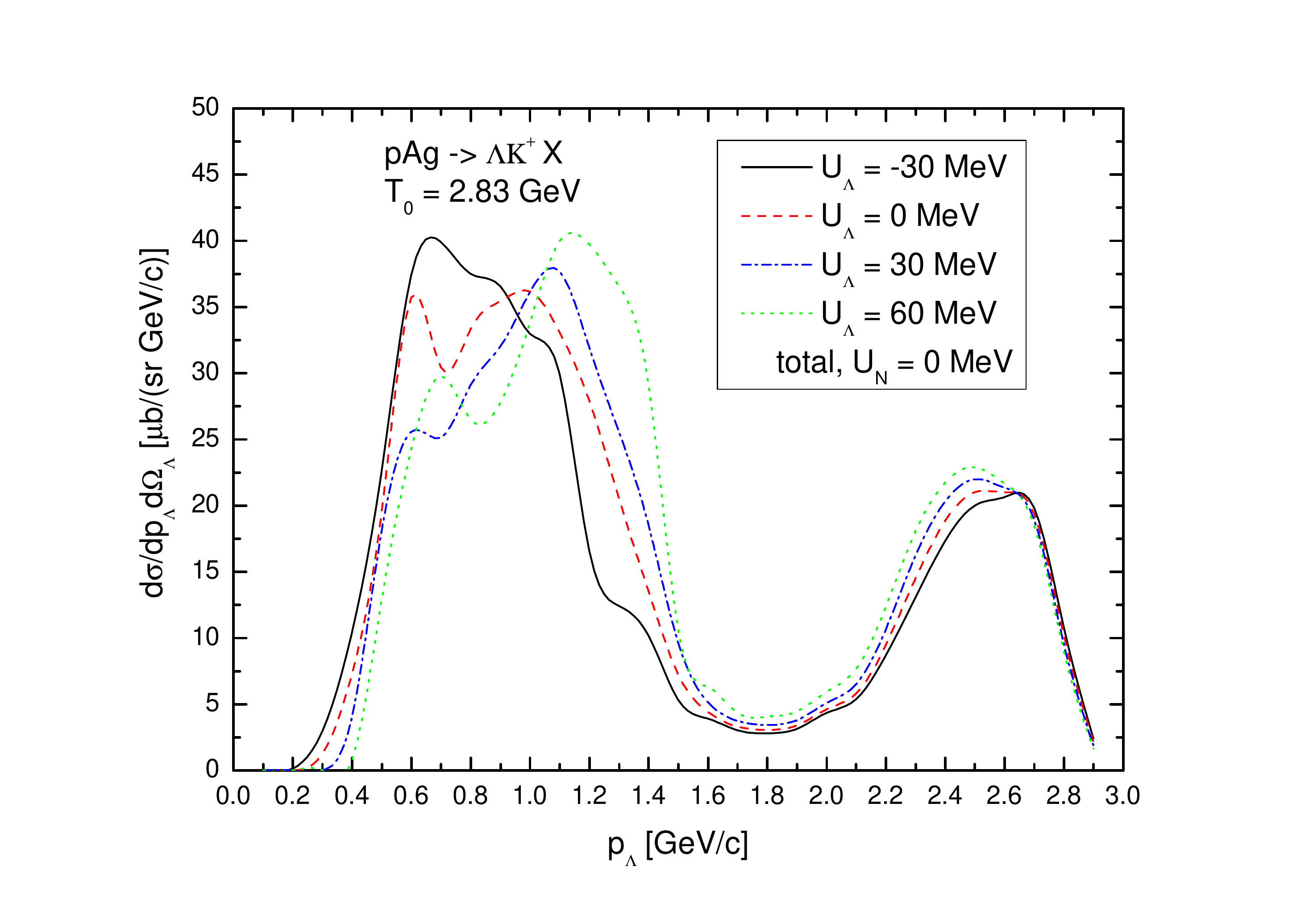}
\vspace*{-2mm} \caption{Differential cross section for the production of $\Lambda$ hyperons
in coincidence with the $K^+$ mesons from primary plus secondary channels
in the ANKE acceptance window as a function of lambda momentum
in the interaction of protons of energy of 2.83 GeV with Ag target nucleus for effective scalar
$\Lambda$ potentials at saturation density
$U_{\Lambda}=-30$ MeV (solid line), $U_{\Lambda}=0$ MeV (dashed line),
$U_{\Lambda}=30$ MeV (dotted-dashed line) and $U_{\Lambda}=60$ MeV (dotted line).}
\label{void}
\end{center}
\end{figure}
\begin{figure}[!h]
\begin{center}
\includegraphics[width=12.0cm]{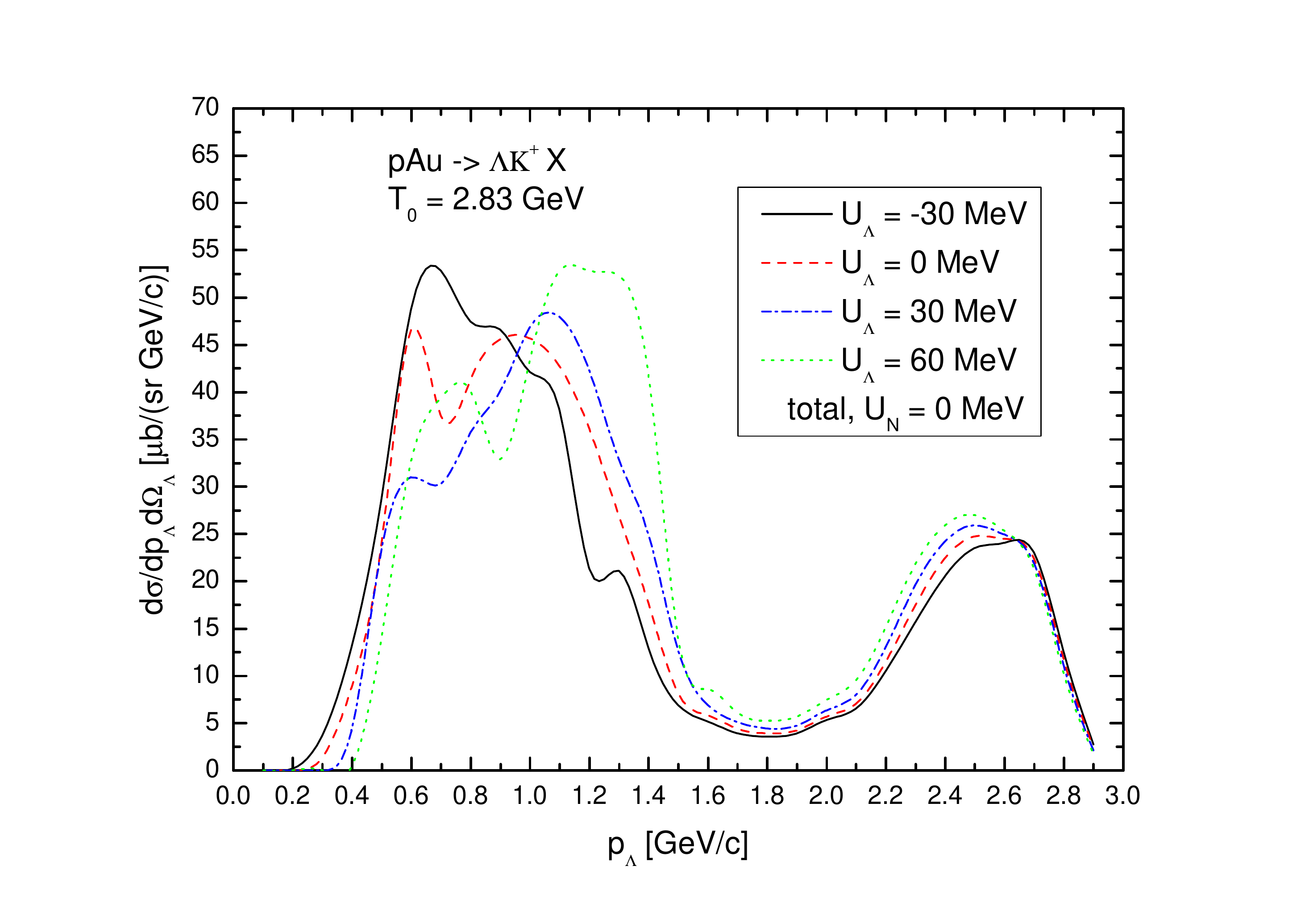}
\vspace*{-2mm} \caption{Differential cross section for the production of $\Lambda$ hyperons
in coincidence with the $K^+$ mesons from primary plus secondary channels
in the ANKE acceptance window as a function of lambda momentum
in the interaction of protons of energy of 2.83 GeV with Au target nucleus for effective scalar
$\Lambda$ potentials at saturation density
$U_{\Lambda}=-30$ MeV (solid line), $U_{\Lambda}=0$ MeV (dashed line),
$U_{\Lambda}=30$ MeV (dotted-dashed line) and $U_{\Lambda}=60$ MeV (dotted line).}
\label{void}
\end{center}
\end{figure}
\begin{figure}[!h]
\begin{center}
\includegraphics[width=12.0cm]{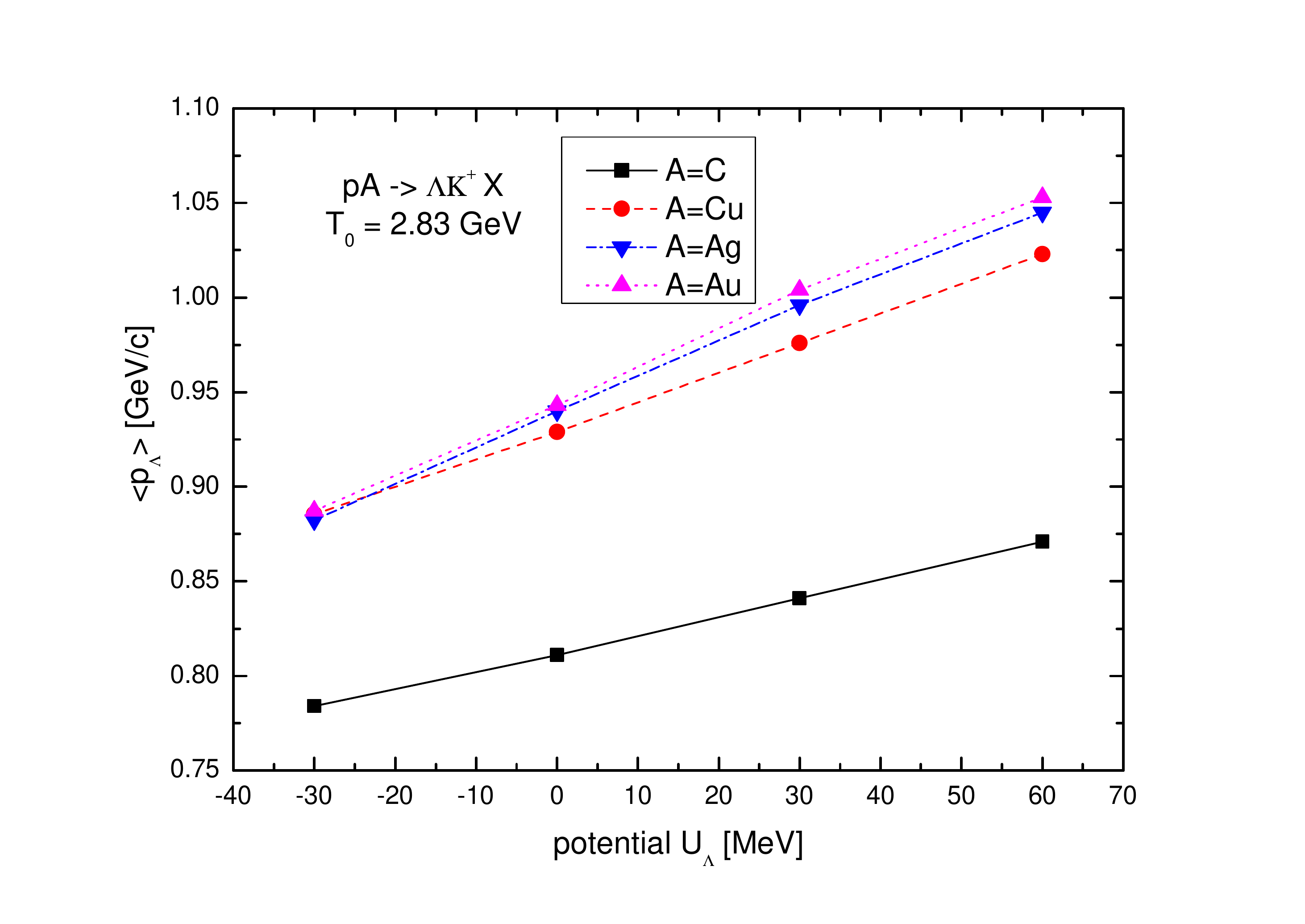}
\vspace*{-2mm} \caption{Average momenta of $\Lambda$ hyperons in the low-momentum
parts of their spectra, shown in Figs.~7--10, from $p$C, $p$Cu, $p$Ag and $p$Au interactions
at an incident energy of 2.83 GeV as functions of the effective scalar $\Lambda$ potential $U_{\Lambda}$
at normal nuclear density. The lines are to guide the eye.}
\label{void}
\end{center}
\end{figure}

\section*{3 Results}

\hspace{1.5cm} First, we consider the differential $\Lambda$ production cross sections
in the ANKE acceptance window from the one-step, two-step and one- plus two-step creation
mechanisms in $p$C and $p$Au reactions at 2.83 GeV beam energy, calculated on the basis of
Eqs.~(46) and (80), in the scenario for the $\Lambda$ effective scalar potential depth
$U_{\Lambda}=-30$ MeV.
These cross sections are presented in Figs.~5 and 6, respectively.
The secondary ${\Lambda}K^+$ production processes (49), (50) with a pion in an
intermediate state are important compared to the primary processes (1)--(6) and the secondary
process (72) associated with the production of $\Lambda$s via the vacuum decay of intermediate
$\Sigma^0$ hyperons, in the chosen kinematics at laboratory lambda momenta $\le$ 1.8 GeV/c, for both
target nuclei. The dominance here is substantially more pronounced for the ${\pi}N \to {\Lambda}K^+$ channels.
At higher $\Lambda$ momenta around 2.4 and 2.7 GeV/c, however, the two-step with
intermediate $\Sigma^0$ hyperons and one-step creation mechanisms are, respectively, dominant.
Evidently, the dependence of the considered coincident $\Lambda$ spectrum on the $\Lambda$ effective scalar
potential should exist mainly at low lambda momenta (cf. Figs.~7--10). This means that
the secondary pion--nucleon production processes have to be accounted for in the analysis of the data on
${\Lambda}K^+$ pair creation in $pA$ collisions obtained in the ANKE experiment with the aim of extracting
the $\Lambda$--nuclear potential. The relative roles of the individual
${\Lambda}K^+$ production channels for other considered options for this potential at saturation
density and target nuclei are similar to those, illustrated in Figs.~5 and 6 for a depth
$U_{\Lambda}=-30$ MeV for C and Au targets. It is clearly seen from Figs.~5 and 6 that the two-bump
structure in the results (and in those shown in Figs.~7--10) is mainly
caused by the two-step (lower bump) and one-step (higher bump) ${\Lambda}K^+$ production mechanisms.

   In Figs.~7--10 we show the results of our calculations following Eqs.~(46) and (80) for the
overall differential cross sections for the production of $\Lambda$ hyperons in coincidence with the
$K^+$ mesons  on C, Cu, Ag and Au target nuclei in the kinematical conditions of the ANKE experiment. These were
obtained for the incident energy of 2.83 GeV, considering four options for the effective scalar
hyperon potential $U_{\Lambda}$ at normal nuclear matter density, as indicated in the insets.
These cross sections are appreciably sensitive to the $\Lambda$ potential at
momenta less than 1.8 GeV/c for all  target nuclei considered; namely, their strengths shift to higher
momenta with increasing $\Lambda$ potential $U_{\Lambda}$ up to 60 MeV. The sensitivity of the strength
of the low-momentum part of the $\Lambda$ spectrum on the scalar $\Lambda$ potential $U_{\Lambda}$, shown in
Figs.~(7)--(10), can be exploited to infer the momentum dependence of
this potential from the direct comparison of the shapes of the calculated $\Lambda$ hyperon
differential distributions with that determined in the ANKE experiment by putting the data in $\Lambda$
momentum bins. As a less differential and an additional measure for the correlation between the
above strength and the $\Lambda$--nuclear potential $U_{\Lambda}$, one can use the average momentum
$<p_{\Lambda}>$, defined as
$<p_{\Lambda}>=\int\limits_{0.1~{\rm GeV/c}}^{1.8~{\rm GeV/c}}p_{\Lambda}dp_{\Lambda}
d\sigma/dp_{\Lambda}d{\bf \Omega}_{\Lambda}/\int\limits_{0.1~{\rm GeV/c}}^{1.8~{\rm GeV/c}}dp_{\Lambda}
d\sigma/dp_{\Lambda}d{\bf \Omega}_{\Lambda}$, where $d\sigma/dp_{\Lambda}d{\bf \Omega}_{\Lambda}$ are the
$\Lambda$ differential cross sections presented in Figs.~7--10.
The use of this quantity has the advantage of significantly decreasing  the uncertainties of absolute
normalization of both the model calculations and the experimental data. The average momentum $<p_{\Lambda}>$
as a function of potential $U_{\Lambda}$ is plotted in Fig.~11.
It is seen that the carbon nucleus is not optimal for determining this potential.
The heavy silver and gold target nuclei show the highest sensitivity to it.
Thus, for example, for the gold nucleus the difference between the mean momenta $<p_{\Lambda}>$ corresponding to the $\Lambda$ potential at saturation density $U_{\Lambda}=-30$ MeV and $U_{\Lambda}=60$ MeV, is
166 MeV/c, whereas the same difference for the carbon target nucleus is only 87 MeV/c
\footnote{$^)$The analogous difference between the average momenta
corresponding to the high-momentum region of 1.8--2.9 GeV/c of the $\Lambda$ spectrum of the  Au nucleus,
as well as to the same potentials $U_{\Lambda}=-30$ MeV and $U_{\Lambda}=60$ MeV, amounts, as
 our calculations show, to 42 MeV/c. This demonstrates the very moderate sensitivity of the
considered momentum distributions to the adopted $\Lambda$ in-medium modification scenarios at high momenta of interest also. }$^)$.
Therefore,
a comparison of the above results with the experimentally determined average momentum in the low-momentum part
of the $\Lambda$ spectrum of the heavy target nuclei under consideration
will also allow one to deduce the effective scalar potential $U_{\Lambda}$ in cold nuclear matter
at this momentum
\footnote{$^)$It should be noted that an analogous possibility was recently realized for the
$\omega$ mesons in Ref.~[70]. }$^)$.
Knowing this potential and using Eqs.~(19) and (20), we can
easily recover the single-particle potential $V_{{\Lambda}A}^{\rm SEP}$ at saturation density for
in-medium momentum $p^{\prime}_{\Lambda}$, corresponding to the experimentally determined average
momentum $<p_{\Lambda}>$. Such a data point may also help to discriminate between the existing models of the
$YN$ interaction at finite momenta.

   Thus, we come to the conclusion that the coincident observables considered above can be useful to help
determine the $\Lambda$--nucleus potential at finite momenta in the region $\le$ 1.8 GeV/c, where the theoretical
predictions for it are available.

\section*{4 Conclusions}

\hspace{1.5cm} In this paper we calculated the momentum dependences of the absolute differential cross
sections for the production of $\Lambda$ hyperons in coincidence with the $K^+$ mesons
from $pA$ ($A$=C, Cu, Ag, and Au) collisions
at 2.83 GeV beam energy in the kinematical conditions of the ANKE experiment, performed at COSY,
by considering incoherent primary proton--nucleon, secondary pion--nucleon ${\Lambda}K^+$
production processes and processes associated with the creation of intermediate ${\Sigma^0}K^+$ pairs
in the framework of a nuclear spectral function approach within the different scenarios for the
$\Lambda$ hyperon effective scalar potential. It was found that the shapes of the cross sections are appreciably sensitive to this potential at $\Lambda$ momenta less than 1.8 GeV/c.
This opens a good possibility to determine the above potential here
from  direct comparison of the results presented in this work with the data
from the ANKE-at-COSY experiment. It was also demonstrated that
the two-step pion--nucleon production channels dominate in the low-momentum ${\Lambda}K^+$ creation in the
chosen kinematics and, hence, they should be taken into account in the analysis of these data with the
purpose of getting definite information on the $\Lambda$ nuclear potential at finite momenta.
\\
\\
{\bf Acknowledgments}
\\
The authors gratefully acknowledge A. Polyanskiy
for his interest in this work, which has been partially financed by the Ministry of Education
and Science of the Russian Federation.
\\

\end{document}